\documentclass{aa}  
\usepackage{dcolumn}
\usepackage{graphicx}
\usepackage{txfonts}
\newcommand{\swift}{{\it Swift}}
\newcommand{\xmm}{XMM-Newton}
\newcommand{\nustar}{NuSTAR}
\newcommand{\iras}{{IRAS\,23226-3843}}
\newcommand{\plm}{$\pm$}
\newcommand{\rb}[1]{\raisebox{1.5ex}[-1.5ex]{#1}}

\makeatletter
 \def\hlinewd#1{%
   \noalign{\ifnum0=`}\fi\hrule \@height #1 \futurelet
    \reserved@a\@xhline}
\makeatother

\newcommand{\kms}{km\,s$^{-1}$}
\newcolumntype{d}{D{.}{.}{-1}}

\begin{document} 

    \title{Optical and X-ray discovery
of the changing-look AGN \iras{}
showing extremely broad and double-peaked Balmer profiles}

   \author{W. Kollatschny \inst{1}, 
           D. Grupe \inst{2},
           M.L. Parker \inst{3}, 
           M. W. Ochmann \inst{1},
           N. Schartel \inst{3},
           E. Herwig \inst{1}, 
           S. Komossa \inst{4},  
           E. Romero-Colmenero \inst{5},
           M. Santos-Lleo \inst{3} 
         }

   \institute{Institut f\"ur Astrophysik, Universit\"at G\"ottingen,
              Friedrich-Hund Platz 1, D-37077 G\"ottingen, Germany\\
              \email{wkollat@astro.physik.uni-goettingen.de}
         \and
         Department of Physics, Earth Sciences, and Space System Engineering, Morehead State University,
          Morehead, KY 40351, USA
         \and
         XMM-Newton Science Operations Centre, ESA, Villafranca del Casuntilo,
         Apartado 78, 28691 Villanueva de la Ca{\~nada}, Spain
         \and
         Max-Planck-Insitut f\"ur Radioastronomie, Auf dem H\"ugel 69,
          D-53121 Bonn, Germany
          \and 
          South African Astronomical Observatory, P.O. Box 9, Observatory 7935, Cape Town, South Africa
}

   \date{Received March 6, 2020; Accepted March 28, 2020}
   \authorrunning{Kollatschny et al.}
   \titlerunning{Variability in \iras}

 
  \abstract
   {}
   {We detected a very strong X-ray decline in the galaxy \iras{} within the XMM-Newton slew
survey in 2017. Subsequently, we carried out multi-band follow-up studies to investigate this 
fading galaxy in more detail.} 
   {We took deep follow-up \swift{}, XMM-Newton, and NuSTAR observations 
in combination with optical SALT spectra of \iras{} in 2017. In addition,
we reinspected optical, UV, and X-ray data that were taken in the past.}
   {\iras{} decreased in X-rays by a factor of more than 30
with respect to ROSAT and \swift{} data taken 10 to 27 years before.
The broadband \xmm/\nustar\ spectrum is power-law dominated, with a contribution
from photoionized emission from cold gas, likely the outer accretion disk or torus.
The optical continuum decreased by 60\%\ 
and the Balmer line intensities decreased
by 50\%\  between 1999 and 2017.
The optical Seyfert spectral type changed simultaneously
with the X-ray flux from a clear broad-line Seyfert 1 type in 1999
to a Seyfert 1.9 type in 2017. 
The Balmer line profiles in \iras{} are extremely broad. The profiles during the
minimum state indicate that they originate in an accretion disk.
The unusual flat Balmer decrement H$\alpha$/ H$\beta$ with a value of 2 indicates a very
high hydrogen density of n$_{H} > $ 10$^{11}$\,cm$^{-3}$ at the center of the accretion disk.
\iras{} shows unusually strong FeII blends with respect to the broad line widths,
in contrast to what is known from Eigenvector 1 studies.
        } 
{}

   \maketitle
%

\section{Introduction}

It is generally known that Seyfert 1 galaxies are variable in the optical
and in X-ray continua on timescales of hours to decades. Several
active galactic nuclei (AGN) have shown variability amplitudes
by a factor of more than 20 in the X-rays
 (e.g., Grupe et al.\citealt{grupe01},\citealt{grupe10}).
AGN that show 
extreme X-ray flux variations in combination with
X-ray spectral variations, that is,\ when a
Compton-thick AGN becomes Compton-thin and vice versa, are called
changing-look AGN (e.g., Guainazzi\citealt{guainazzi02}). 
By analogy, optical changing-look AGN exhibit transitions from type 1 to type 2 and vice
versa. In this case, the optical spectral classification can change as a result of a variation
in the accretion rate, accretion disk instabilities, or a variation in reddening.

To date, a few dozen Seyfert galaxies are known to have changed their optical
spectral type, for example, NGC\,3516 (Collin-Souffrin et al.\citealt{souffrin73}),
NGC\,7603 (Tohline \& Osterbrock\citealt{tohline76}, Kollatschny et al.\citealt{kollatschny00}),
 NGC\,4151 (Penston \& Perez\citealt{penston84}), Fairall\,9 
 (Kollatschny et al.\citealt{kollatschny85}), NGC\,2617 (Shappee
 et al.\citealt{shappee14}), Mrk\,590 (Denney at al.\citealt{denney14}), 
HE 1136-2304  (Parker et al.\citealt{parker16}, Zetzl et al.\citealt{zetzl18},
Kollatschny et al.\citealt{kollatschny18}), and
1ES\,1927+654 (Trakhtenbrot et al.\citealt{trakhtenbrot19}, and references therein).
Further recent findings are based on spectral
variations detected by means of the Sloan Digital Sky Survey (SDSS)
(e.g., Komossa et al.\citealt{komossa08}, LaMassa et al.\citealt{lamassa15}, 
Rumbaugh et al.\citealt{rumbaugh18}, MacLeod et al.\citealt{macleod19},),
the Catalina Real-time Transient Survey (Graham et al.\citealt{graham20}), 
or the Wide-field Infrared Survey Explorer (Stern et al.\citealt{stern18}). 
In most of these recent findings, only a few optical spectra
of the individual SDSS galaxies 
have been secured to prove their changing-look character.
Other studies tried to detect correlations of the optical variability amplitudes 
with spectral properties{\bfseries,} such as emission line widths (Kollatschny
et al.\citealt{kollatschny06}).

\begin{figure}
\centering
\includegraphics[width=8.5cm,angle=0]{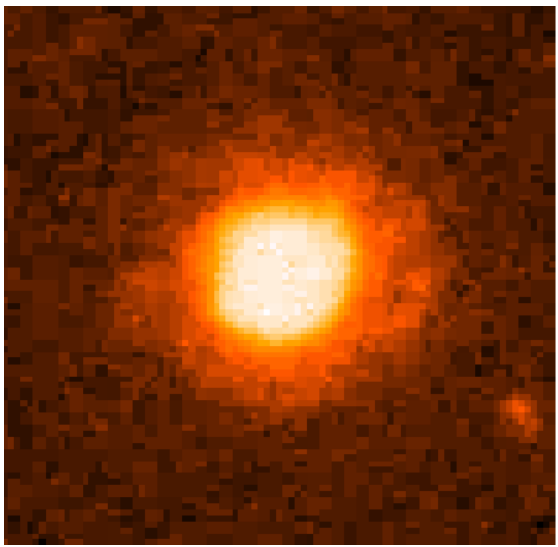}
\caption{6dF B-band image of the S0-type galaxy \iras{}. Scale: 1 x 1 arcmin. North is
to the top, east to the left.
  }
\label{6dF_UKST_B_IRAS23252.pdf}
\end{figure}

Initially, \iras{} was listed as a bright X-ray source in the RASS catalog
(e.g., Grupe et al.\citealt{grupe01}).
Other designations for \iras{} are CTS~B11.01, 2MASX~J23252420-3826492,
6dF J2325242-382649, and HE~2322-3843.
It was included in the NRAO VLA Sky Survey (NVSS)  as a radio source 
with a flux density level of 4.3\,mJy (Condon et al.\citealt{condon98}).
\iras{} was classified as Seyfert type 1
in 1991 (Allen et al.\citealt{allen91}) based on its optical spectrum. 
The galaxy is of morphology type SO (Loveday\citealt{loveday96}).
Figure~\ref{6dF_UKST_B_IRAS23252.pdf}
displays the 6df B-band image of \iras{}
(scale: 1 x 1 arcmin corresponding to 41 x 41 kpc) (Jones\citealt{jones04}).
An apparent magnitude of 13.38$\pm$0.10 was measured for 
this object in the R band in the Las Campanas Redshift Survey (1991-1996)
(Shectman et al.\citealt{shectman96}). This corresponds to an absolute magnitude
of M$_{R}$=$-22.33\pm0.10$~mag.

We detected a very strong X-ray decrease in \iras{} within the XMM-Newton slew
survey in 2017. Thereupon, 
we carried out optical and X-ray follow-up studies to investigate 
this galaxy in more detail and to study its variability behavior. 
Extreme cases of Seyfert-type changes like those in \iras{} provide us with tight constraints
on the accretion physics and accretion disk changes on short timescales.

Throughout this paper, we assume a $\Lambda$CDM cosmology
with a Hubble constant of H$_0$~=~73~km s$^{-1}$ Mpc$^{-1}$, 
 $\Omega_{\Lambda}$=0.73, and $\Omega_{\rm M}$=0.27.
With a redshift of {\bfseries $z=0.0359,$} this results in a luminosity distance
of $D_{\rm L}$ = 144 Mpc.

\section{Observations and data reduction}
\iras{} ($\alpha_{2000}$ = 23h 25m 24.2s,
$\delta$ = $-38^{\circ}$ 26$^{'}$ 49.2$^{''}$)
was found to be a variable X-ray source by the XMM-Newton slew
survey in April, 2017.
The 0.2--2 keV flux decreased by a factor of about 10 in comparison
to ROSAT all-sky survey data in 1990. The ROSAT flux in 1990 amounted to
$1.85\pm0.18\times10^{-11}\,cm^{-2}$\,s$^{-1}$ (0.1-2.4 keV) 
(Mahony et al.\citealt{mahony10}).
We started a variability campaign with \swift\  after detecting the decreasing X-ray flux
in 2017 April. In addition, we carried out follow-up \xmm\ and \nustar\ observations as part of 
a ToO program to study extreme variable X-ray sources. Furthermore, we took optical spectra
of \iras{} with SALT to investigate the optical spectral variations.

\subsection{\bf X-ray, UV, and optical continuum observations with \swift{}}
 
After the discovery of the decreasing X-ray flux in 2017 April,
 we started monitoring \iras{} 
with \swift\ 
(Gehrels et al.\citealt{gehrels04}) in X-rays and the UV/optical. All \swift\ 
observing dates and exposure times are listed in Table\,\ref{swiftlog}. 
In addition, \iras{} has been observed before by \swift\ during three epochs in
2007, 2008-2009, and 2013. For comparison purposes,
we list these observations in all the tables relevant to \swift\ data 
(see Tables\, \ref{swiftlog}, \ref{swiftmag}, and \ref{swiftdata}).

All X-ray observations with the \swift\ X-ray Telescope (XRT, Burrows et al.\citealt{burrows05}) were
performed in photon-counting mode (pc-mode, Hill et al.\citealt{hill04}). 
During the high state at the beginning of the observations in 2007-2013, 
source counts were collected in a circular region with a radius of 30 pixels (equivalent to 70$^{''}$) and
background counts in a nearby source-free circular region with a radius of 90 pixels (equal to 210${''}$).
For the low-state observations in 2016-2017, we used  47$^{''}$ (20 pixel) and 235$^{''}$ 
(100 pixel) extraction radii for the source and background counts, respectively.
Spectra were extracted with the FTOOL {\it XSELECT}. An auxiliary response file (ARF) was created
for each observation using  {\it xrtmkarf}. We applied the \swift\ XRT response file {\it swxpc0to12s6\_20130101v014.rmf}.  

Most spectra were rebinned to have at least 20 counts per bin using {\it grppha}.
For low-state spectra, however,  the number of counts was too low to allow $\chi^2$
statistics. These data were fit by Cash statistics (Cash\citealt{cash79}) as
indicated in Table\,\ref{swiftdata}.
The spectral analysis was performed in {\it XSPEC}
(Arnaud\citealt{arnaud96}).
All spectra were fit with a single power-law
 model with Galactic absorption ($N_{\rm H,gal} = 1.59\times 10^{20}$ cm$^{-2}$;
Kalberla et al. \citealt{kalberla05}). 

Count rates, hardness ratios, and the best-fit values obtained are listed
in Table\,\ref{swiftdata}. 
The hardness ratio is defined as $HR = \frac{H-S}{H+S}$ , where
S = counts(0.3--1.0 keV) and H = counts(1.0--10.0 keV).
In order to determine a background-corrected hardness ratio, we
applied the program  {\it BEHR} by Park et al.\cite{park06}. 

During most observations, the \swift\ UV-Optical Telescope (UVOT, Roming et
al. \citealt{roming05}) observed in all six photometric filters UVW2 (1928
\AA{}), UVM2 (2246 \AA{}),  UVW1 (2600 \AA{}), u (3465\AA{}), b (4392 \AA{}),
and v (5468 \AA{}). Before we analyzed the data, all snapshots in one segment were combined with the
UVOT tool {\it uvotimsum}.
The flux densities and magnitudes in each filter were determined by the tool {\it uvotsource} using 
the count rate conversion and 
calibration, as described in  Poole et al.\cite{poole08} and  Breeveld et al.\cite{breeveld10}. 
Source counts were extracted in a circle with a radius of $3^{''}$ and
background counts in a nearby source-free region with a radius of
20$^{''}$. The smaller source extraction radius of 3$^{''}$ was corrected for by setting the UVOT task
{\it uvotsource} parameter {\it apercorr = curveofgrowth}. 
The UVOT fluxes listed in Table\ref{swiftmag} are not corrected
 for Galactic reddening.
The reddening value in the direction of \iras{} is
E$_\text{B-V}$ = 0.025,  deduced from
 the Schlafly \& Finkbeiner\cite{schlafly11} recalibration of the 
Schlegel et al.\cite{schlegel98} infrared-based dust map.
Applying Eq. (2) in Roming et al.\cite{roming05}, 
who used the standard reddening correction curves by Cardelli et al.\cite{cardelli89}, we calculated the following 
 magnitude corrections:
v$_\text{corr}$ = 0.083, b$_\text{corr}$ = 0.109, u$_\text{corr}$ = 0.136,
UVW1$_\text{corr}$ = 0.177, UVM2$_\text{corr}$ = 0.244, and UVW2$_\text{corr}$ = 0.205.
For all \swift\, UVOT magnitudes used in this publication, we adopted the Vega magnitude system.

\subsection{{\bf X-ray observations with \xmm\ and \nustar\ }}
The \xmm\ and \nustar\ observations were taken as part of a ToO program (Proposal ID 08205301, PI Schartel) aimed
at catching AGN in deep minimum or maximum states (see, e.g., Parker et al.\citealt{parker16},
\citealt{parker19}).
The observing program consisted of one initial \xmm\ snapshot, a
follow-up long \xmm/\nustar\ exposure, and a final follow-up snapshot with \xmm . The details of these
observations are given in Table~\ref{table:xray_observations}.

\begin{table*}
\centering
\caption{Details of the \xmm\ and \nustar\ X-ray observations. Exposure times are the
final clean exposures after filtering for background flares.}
\label{table:xray_observations}
\begin{tabular}{l c c c c}
\hline 
\noalign{\smallskip}
Mission & Obs. ID &  Julian date &  Start date & Exp. time\\
        &         & 2\,400\,000+ &             &   [ks]   \\    
\hline
\xmm\     & 0760020101  &  57875  & 2017-05-02    & 12 \\
\xmm\     & 0760020401  &  57915  & 2017-06-11    & 76 \\
\nustar\  & 80101001002 &  57915  & 2017-06-11    & 96 \\
\xmm\     & 0760020501  &  58071  & 2017-11-14    & 9  \\
\hline
\end{tabular}
\end{table*}

We reduced the \xmm\ data using the \xmm\ Science Analysis Software (SAS) version 16.1.0. We extracted EPIC-pn
and EPIC-MOS spectra for each observation using the \emph{epproc} and \emph{emproc} tools, and filtered
for background flares. We took source photons from $30^{\prime\prime}$ circular regions centered on the source
position and background photons from large circular regions on the same chip where possible and avoided
contaminating sources and the high copper background region of the pn. There was almost no variability during
the observations, therefore we focused on analyzing the spectra.

The three \xmm\ spectra are very similar. Preliminary fitting of the spectra with a power law showed
that the spectral indices of the observations are constant, the residuals do not change between
observations, and there is only a mild difference in normalization. Because the two short observations
do not show major differences and the signal-to-noise ratio (S/N) is too low to reveal smaller changes,
we combined the spectra from the three observations into a single spectrum for the remainder
of this work. We also combined the spectra from MOS1 and MOS2 into a single combined MOS spectrum.

We reduced the \nustar\ data using the \nustar\ Data Analysis Software (NuSTARDAS) version 1.8.0, and
extracted source photons from a 30$^{\prime\prime}$ circular region centered on the source position.
We extracted background photons from a larger 80$^{\prime\prime}$ circular region on the same chip and avoided
contaminating sources.

We binned all spectra to an S/N of 6, and to oversample the spectral resolution
by a factor of 3. We fit the X-ray spectra
in \textsc{xspec} version 12.9.1p.

\subsection{{\bf Optical spectroscopy with the SALT telescope}}
\begin{table}
\centering
\caption{Log of spectroscopic observations of \iras{}
at the Cerro Tololo Inter-American Observatory in 1999 and
with the SALT telescope in 2017.
Listed are the Julian date, the UT date, and the exposure time.}
\begin{tabular}{c c c l}
\hline 
\noalign{\smallskip}
Julian date &  & Exp. time & comments\\
2\,400\,000+&  \rb{UT date}       &  [s] &  \\
\hline 
51350         &       1999-06-21      &      600  & clouds  \\
57883         &       2017-05-10      &      900  & clear, 1\farcs{}5    \\
57916         &       2017-06-12      &      900  & clear, 2\farcs{}0   \\ 
\hline 
\vspace{-.7cm}
\end{tabular}
\label{saltlog}
\end{table}

An early optical spectrum of \iras{} was taken with the 4m Blanco telescope at
the Cerro Tololo Inter-American Observatory  in Chile on 1999 June 21 under nonphotometric conditions
(Grupe et al.\citealt{grupe04})\footnote{The optical spectrum of \iras{} 
presented in that paper has erroneously been mixed up.}. The R-C spectrograph was attached to the
telescope, and a slit width of 2$^{\prime\prime}$ was used oriented in east-west direction.

We took a new optical spectrum of \iras{}
with the 10\,m Southern African Large Telescope (SALT)
after the detection of the X-ray variability on 2017 May 10.
A second optical spectrum was secured
simultaneously to the deep X-ray observation of XMM-Newton
on 2017 June 12.
 The log of our optical spectroscopic observations
with SALT is given in Table~\ref{saltlog}.

The spectroscopic observations with the SALT telescope were taken under
identical instrumental conditions with the Robert Stobie Spectrograph
using the PG0900 grating.
The slit width was fixed to
2\arcsec\hspace*{-1ex}.\hspace*{0.3ex}0 projected onto the
sky at an optimized position angle to minimize differential refraction.
The spectra were taken with an exposure time of 
15 minutes (see Table~\ref{saltlog}).
Seeing full width at half maximum (FWHM) values 
were 1.5 to 2\arcsec.

We covered the wavelength range from
4330 to 7376~\AA\  at a spectral resolution of 6.5~\AA\ .
The observed wavelength range corresponds to 
a wavelength range from 4180 to 7120~\AA\ in the rest frame of the galaxy. 
 There are two gaps in the spectrum caused by the gaps between the three CCDs:
one between the blue and the central CCD chip, and one between the
central and red CCD chip covering the wavelength ranges
5347--5416~\AA\  and 6386--6454~\AA\ (5162--5228~\AA\ and 6165--6230~\AA\
in the rest frame).
All spectra were wavelength corrected to the rest frame of the galaxy
{\bfseries ($z=0.0359$)}. 

In addition to the galaxy spectra, we also observed necessary flat-field and
ThAr and Xe arc frames, as well as spectrophotometric
standard stars for flux calibration (LTT1020).
The spatial
resolution per binned pixel was 0\arcsec\hspace*{-1ex}.\hspace*{0.3ex}2534
for our SALT spectra.
We extracted  nine columns from our object spectrum, 
corresponding to 2\arcsec\hspace*{-1ex}.\hspace*{0.3ex}28.
The reduction of the spectra (bias subtraction, cosmic-ray correction,
flat-field correction, 2D wavelength calibration, night sky subtraction, and
flux calibration) was made in a homogeneous way with IRAF reduction
packages (e.g., Kollatschny et al.\citealt{kollatschny01}).  
We obtained S/N values higher than 50 in the continuum.

Great care was taken to ensure high-quality intensity and wavelength
calibrations to keep the intrinsic measurement errors very low, as described in
Kollatschny et al.\cite{kollatschny01,kollatschny03,kollatschny10}.
The AGN spectra (and  our
 calibration star spectra) were not always taken
 under photometric conditions. 
Therefore
we calibrated the spectra to the same absolute
[\ion{O}{iii}]\,$\lambda$5007 flux of
$1.01 \times 10^{-14} \rm erg\,s^{-1}\,cm^{-2}$ 
($9.27 \times 10^{-15} \rm erg\,s^{-1}\,cm^{-2}$ in rest frame) 
taken on 2017 May 10 under clear conditions.
The flux of the narrow emission line [\ion{O}{iii}]\,$\lambda$5007
is considered to be constant on timescales of many years.

\section{Results}

\subsection{\bf X-ray, UV, and optical continuum variations}
\begin{figure*}
\centering
\vspace*{-20mm} 
\includegraphics[width=20.cm,angle=0]{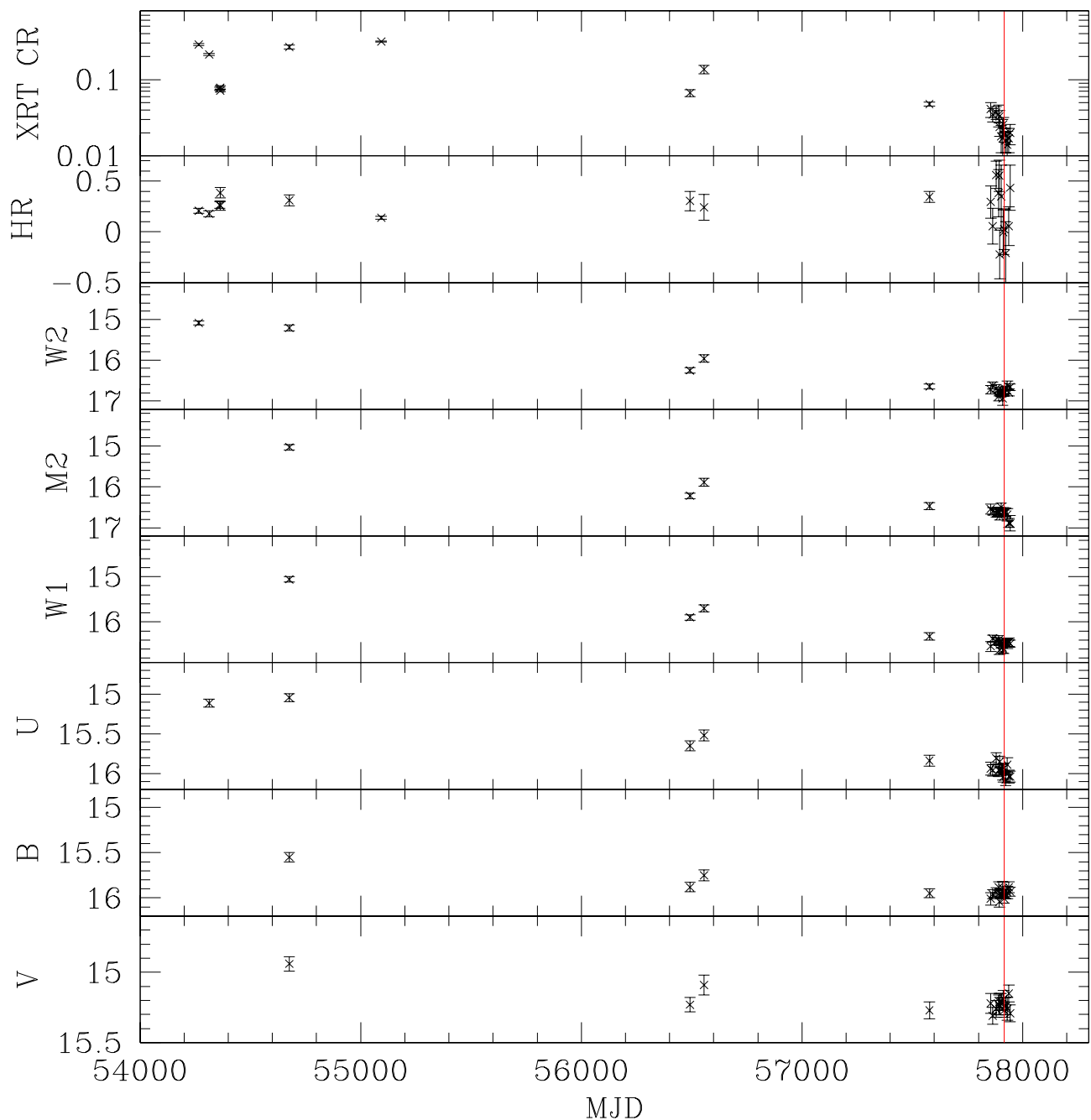}
\vspace*{-50mm} 
\caption{Combined X-ray, UV, and optical light curves
taken with the \swift{} satellite for the years 2007 until 2017. The red line indicates the date of our deep 
XMM-Newton observation on 2017-06-11.
}
\label{fig_dirk_xrt_uvot_lc07_17.pdf}
\end{figure*}

\begin{figure*}
\centering
\vspace*{-20mm} 
\includegraphics[width=20.cm,angle=0]{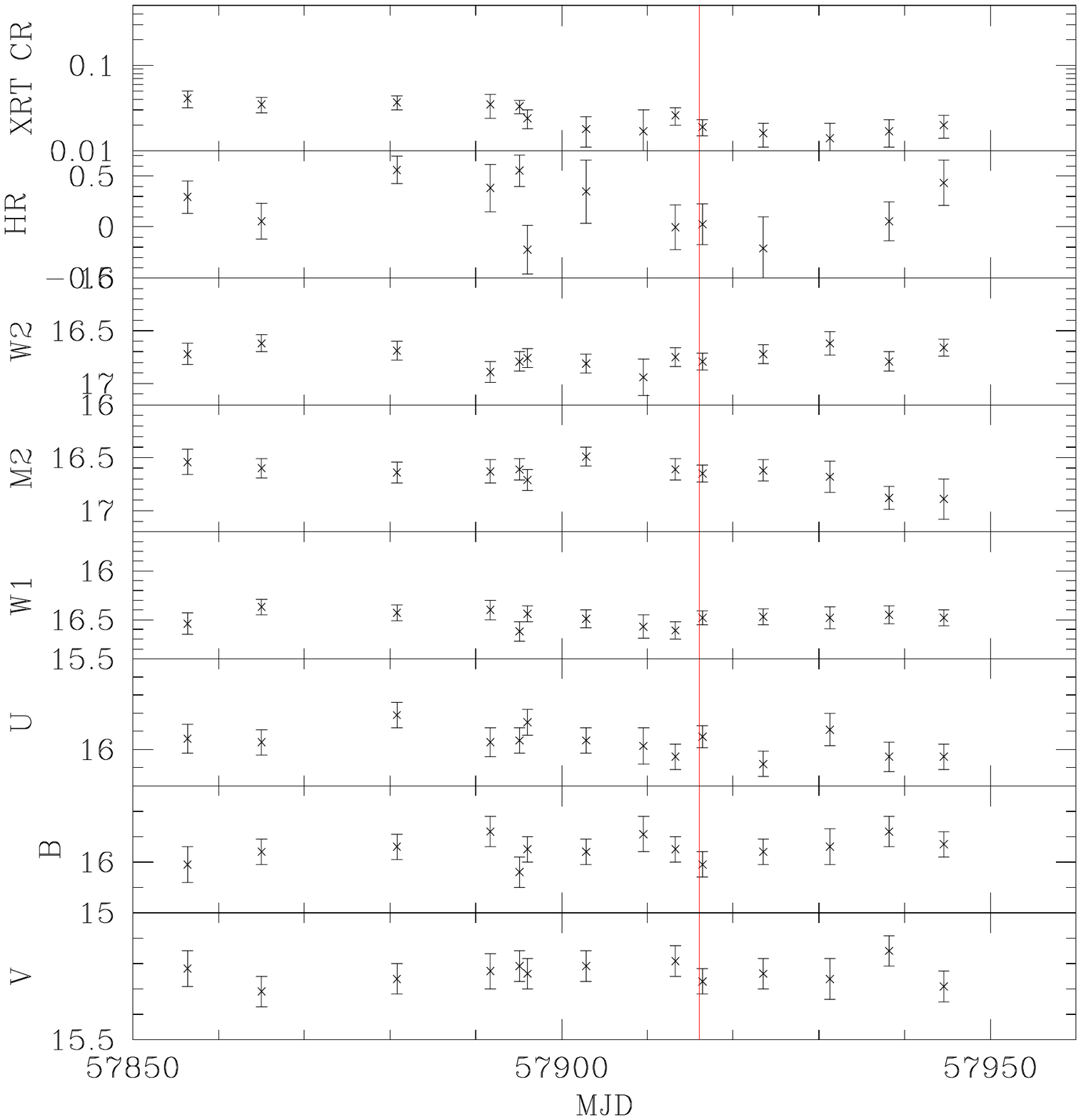}
\vspace*{-50mm} 
\caption{Combined X-ray, UV, and optical light curves
taken with the \swift{} satellite for the year 2017. The red line indicates the date of our deep 
XMM-Newton observation on 2017-06-11.
}
\label{fig_dirk_xrt_uvot_lc2017.pdf}
\end{figure*}

\swift{} 0.3--10 keV, UV, and optical light curves
for the years 2007 to 2017 are shown in
 Figures~\ref{fig_dirk_xrt_uvot_lc07_17.pdf}
and \ref{fig_dirk_xrt_uvot_lc2017.pdf} along with
the X-ray count rates
and the hardness ratios (see Sect. 2.2). All measurements are listed in 
Tables\,\ref{swiftmag} and \ref{swiftdata}.
The X-ray 0.3--10 keV flux and count rate are
clearly variable and decrease by a factor of 35. 
 We also checked whether there is any significant variability in
the hardness ratio and photon index $\Gamma$.
There is no obvious connection with the X-ray flux and count rate
variability and the variability of the hardness ratio. The distribution of
the hardness ratios is almost Gaussian.
Figure~\ref{fig_dirk_xrt_uvot_lc2017.pdf} 
shows the X-ray, UV, and optical \swift{} light curves
for the detailed campaign in 2017. 
The UV and optical \swift{} bands closely follow the X-ray light curve. The
X-ray light curve exhibits the strongest variability amplitudes.
Table~\ref{swiftvarstatistics}
gives the variability statistics based on
the \swift{} continua (XRT, W2, M2, W1,
 U, B, and V). 
 \begin{table}
\centering
\tabcolsep+1.5mm
\caption{Variability statistics based on the \swift{} continua (XRT, W2, M2, W1, U, B, V)
in units of 
10$^{-12}$\,erg\,s$^{-1}$\,cm$^{-2}$\,\AA$^{-1}$ and $10^{-12}$ ergs s$^{-1}$ cm$^{-2}$ for the 0.3--10 keV X-ray data.
}
\begin{tabular}{lrrrccc}
\hline 
\noalign{\smallskip}
    & F$_\text{max}$  & F$_\text{min}$  & R$_\text{max}$  & <F>  &  $\sigma_\text{F}$ &  F$_\text{var}$ \\
\noalign{\smallskip}
\hline 
\noalign{\smallskip}
XRT     &   11.45   &  0.33    &   34.70   &   2.03 & 2.56 &  1.25 $\pm$   0.62     \\
W2      &    8.45   &  1.94    &    4.36   &   2.65 & 2.91 &  1.10 $\pm$   0.41       \\
M2      &   10.14   &  1.84    &    5.51   &   3.02 & 2.07 &  0.68 $\pm$   0.23       \\
W1      &    9.79   &  2.34    &    4.18   &   3.39 & 1.87 &  0.54 $\pm$   0.16       \\
U       &   11.74   &  4.59    &    2.56   &   5.75 & 1.71 &  0.29 $\pm$   0.05       \\
B       &   17.18   & 10.87   &    1.58    &  12.30 & 1.48 &  0.11 $\pm$   0.01       \\
V       &   21.59   & 15.4    &    1.40    &  16.85 & 1.41 &  0.06 $\pm$   0.005  \\
\noalign{\smallskip}    
\hline
\hline
\end{tabular}
\label{swiftvarstatistics}
\end{table}
We indicate the minimum and maximum fluxes F$_{\rm min}$ and F$_{\rm max}$,
peak-to-peak amplitudes R$_{\rm max}$ = F$_{\rm max}$/F$_{\rm min}$, the mean flux $<$F$>$ 
over the period of observations
from JD 54676.78 to 57944.56, 
the standard deviation $\sigma_{\rm F}$,  and the fractional variation 
\begin{align*}
F_{\rm var} = \frac{\sqrt{{\sigma_F}^2 - \Delta^2}}{<F>}~, 
\end{align*}
as defined by Rodr\'\i{}guez-Pascual et al.\cite{rodriguez97}.
The quantity $\Delta^2$ is the mean square value of the uncertainties 
$\Delta_\text{i}$ associated with the fluxes $F_\text{i}$.
The $F_{\rm var}$ uncertainties are defined in Edelson et al.\cite{edelson02}.
The peak-to-peak amplitude and the fractional variation decrease as a function
of wavelength.

\subsection{\bf X-ray spectrum}

\begin{figure*}
\centering
\includegraphics[width=12.1cm,angle=0]{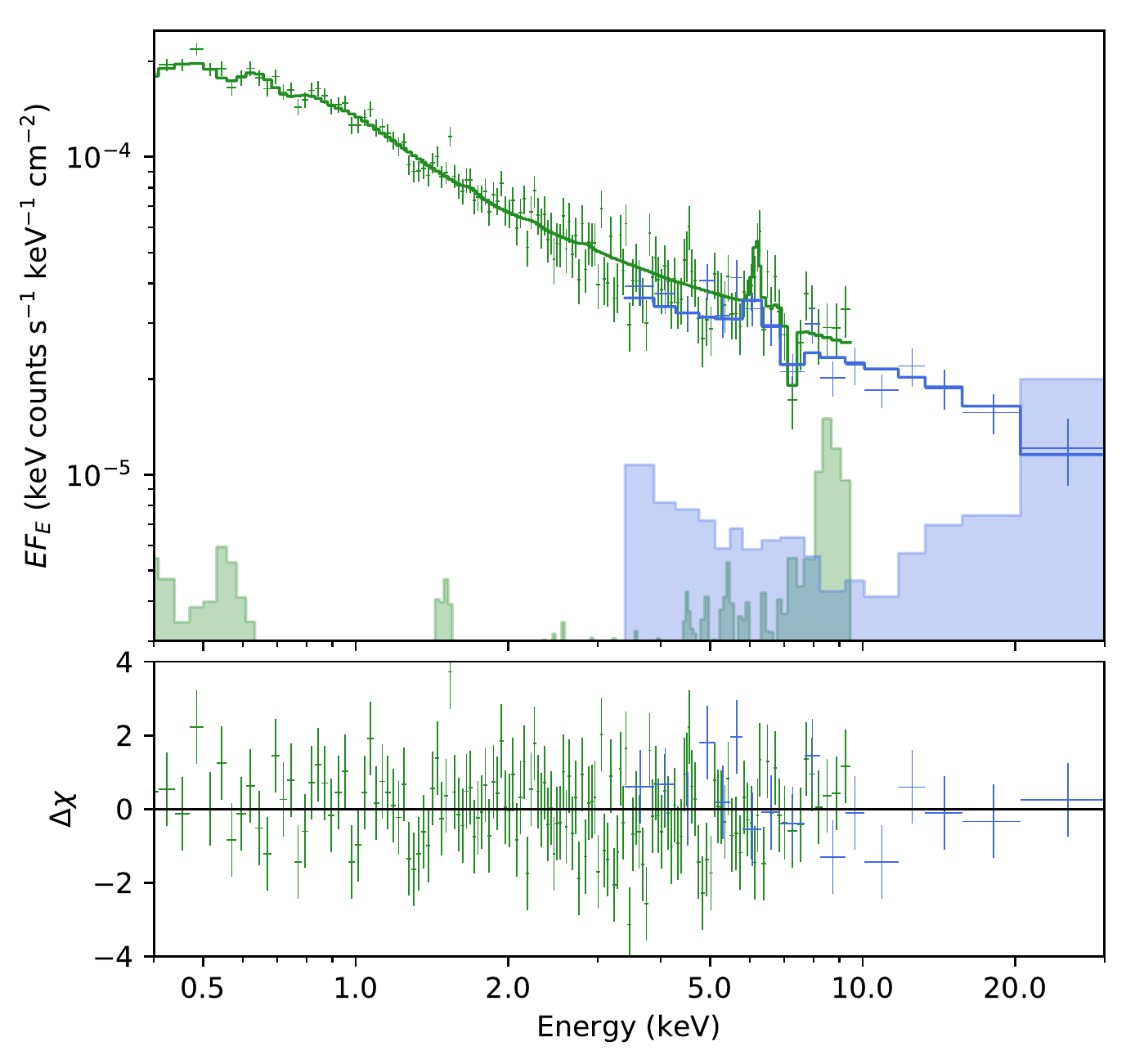}
\caption{XMM-Newton and NuSTAR spectrum of \iras{}. The data are corrected for 
the effective area of the instruments (using the `setplot area' command in \textsc{xspec})
but are not unfolded. Solid lines show the best-fit model, and shaded 
regions show the background spectra. The lower panel shows the residuals to the 
best-fit model. For clarity, we show only the EPIC-pn and \nustar\ data. The two 
 \nustar\ FPM spectra are grouped for clarity in the figure,
 but fitted separately.}
\label{parkerbestfit_data_pn_and_nustar.pdf}
\end{figure*}

\begin{figure*}
\centering
\includegraphics[width=12.1cm,angle=0]{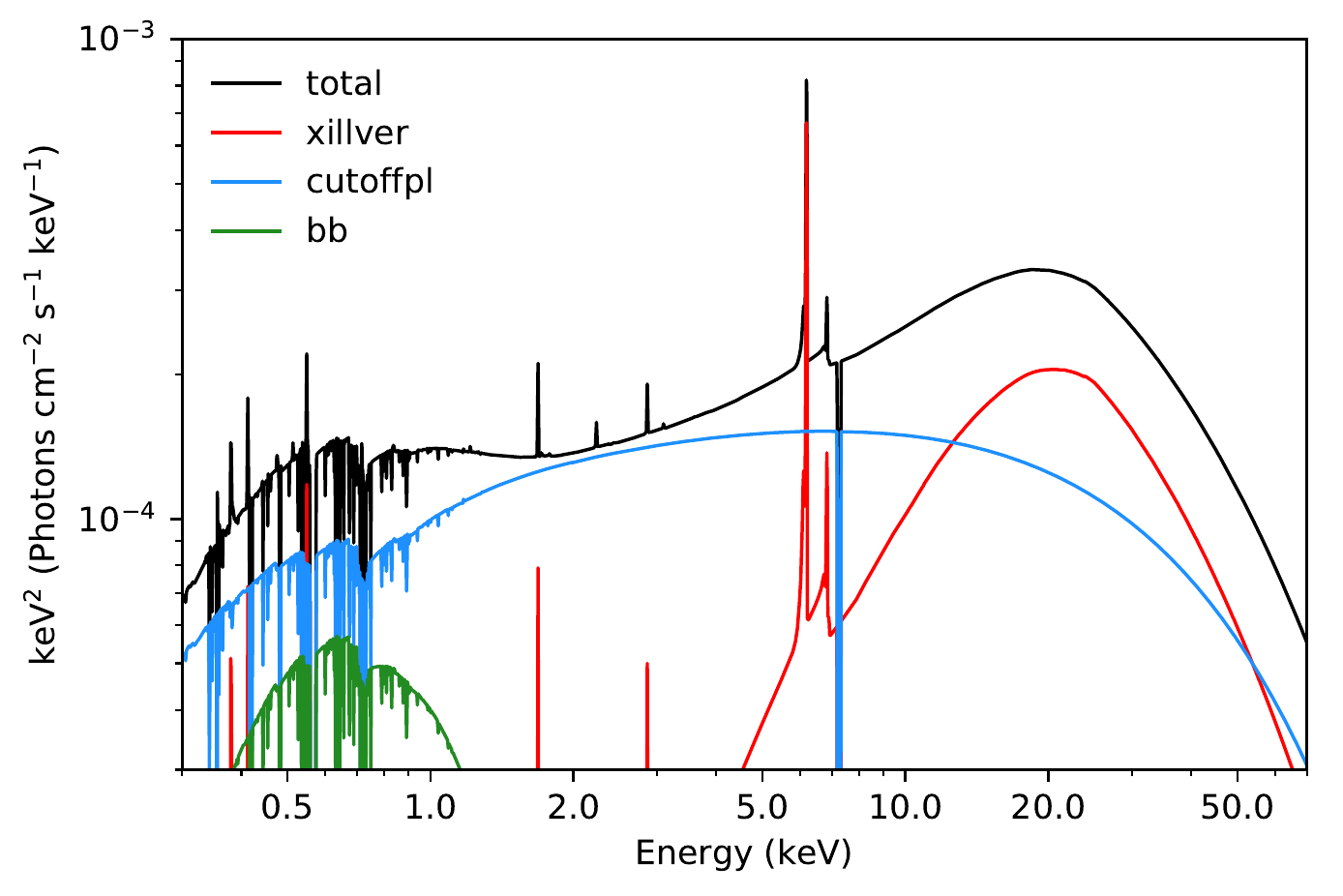}
\caption{Best-fitting broadband spectral model, showing
the different spectral components.}
\label{parkerbestfit_model_pn.pdf}
\end{figure*}

The broadband \xmm/\nustar\ spectrum shows a soft excess, a weak narrow Fe line 
at 6.4~keV, and a possible absorption feature just above 7~keV. We fit the 
spectrum with Galactic absorption (modeled with TBnew; Wilms et al., 2000), a cut-off
 power-law continuum, a phenomenological blackbody soft excess, neutral reflection
modeled with \emph{xillver} (Garcia et al., 2013) a warm absorber modeled 
with \emph{xstar} (Kallman \& Bautista, 2001), and a Gaussian absorption line. We
 also allowed for a constant offset between the different detectors to allow for 
calibration differences and the nonsimultaneous \xmm/\nustar\ exposures. The 
difference between \xmm\ and \nustar\ is relatively large because EPIC 
includes the two shorter intervals when \nustar\ was not observing.
The data are shown in Fig.~\ref{parkerbestfit_data_pn_and_nustar.pdf}, the best-
fit model is shown in Fig.~\ref{parkerbestfit_model_pn.pdf}, and the parameters 
are given in Table~\ref{table:bestfit_data_pn_and_nustar}.
\begin{table}
\caption{Best-fit model parameters for the broadband X-ray spectrum. Normalization
constants are defined relative to the EPIC-pn normalization.}
\label{table:bestfit_data_pn_and_nustar}
\begin{tabular}{l l r}
\hline
Parameter & Value & Description\\
\hline
$E_\mathrm{Gauss}$     &     $7.34_{-0.12}^{+0.07}$  &       Gaussian energy (keV)\\
$N_\mathrm{H}$         &     $(4\pm2)\times10^{20}$  &       Column density (cm$^{-2}$)\\
$\log(\xi)$            &     $0.6^{+0.3}_{-0.5}$     &       Ionization (erg~cm~s$^{-1}$)\\
$\Gamma$               &     $1.75_{-0.04}^{+0.03}$  &       Photon index\\
$E_\mathrm{cut}$       &     $30_{-10}^{+20}$        &       Cut-off energy (keV)\\
$A_\mathrm{Fe}$        &     $<0.8$                  &       Iron abundance\\
$kT$                   &     $0.17\pm0.01$           &       Blackbody temperature (keV)\\
\hline
$C_\mathrm{pn}$        &     $1^*$                   &       EPIC-pn constant\\
$C_\mathrm{MOS}$       &     $1.09\pm0.01$           &       EPIC-MOS constant\\
$C_\mathrm{FPMA}$      &     $0.84\pm0.05$           &       FPMA constant\\
$C_\mathrm{FPMB}$      &     $0.82\pm0.05$           &       FPMB constant\\
\hline
$\chi^2$/dof           &     421/395                  & Fit statistic\\
\hline
\end{tabular}
\end{table}

This model gives a good fit to the data, with no obvious structured residuals. The significance of the 7.3~keV line
is low: although a residual feature is present in all three instruments, it
is weak, and the difference in $\chi^2$ is only 7 for two additional degrees of freedom. If this were real, it
would correspond to an outflow with a velocity of $\sim0.05c$, assuming the line is produced by Fe~\textsc{xxvi}, but
we cannot claim a detection based on the current data.

\subsection{\bf Optical spectra and their variations}
\begin{figure*}
\centering
\includegraphics[width=19.cm,angle=0]{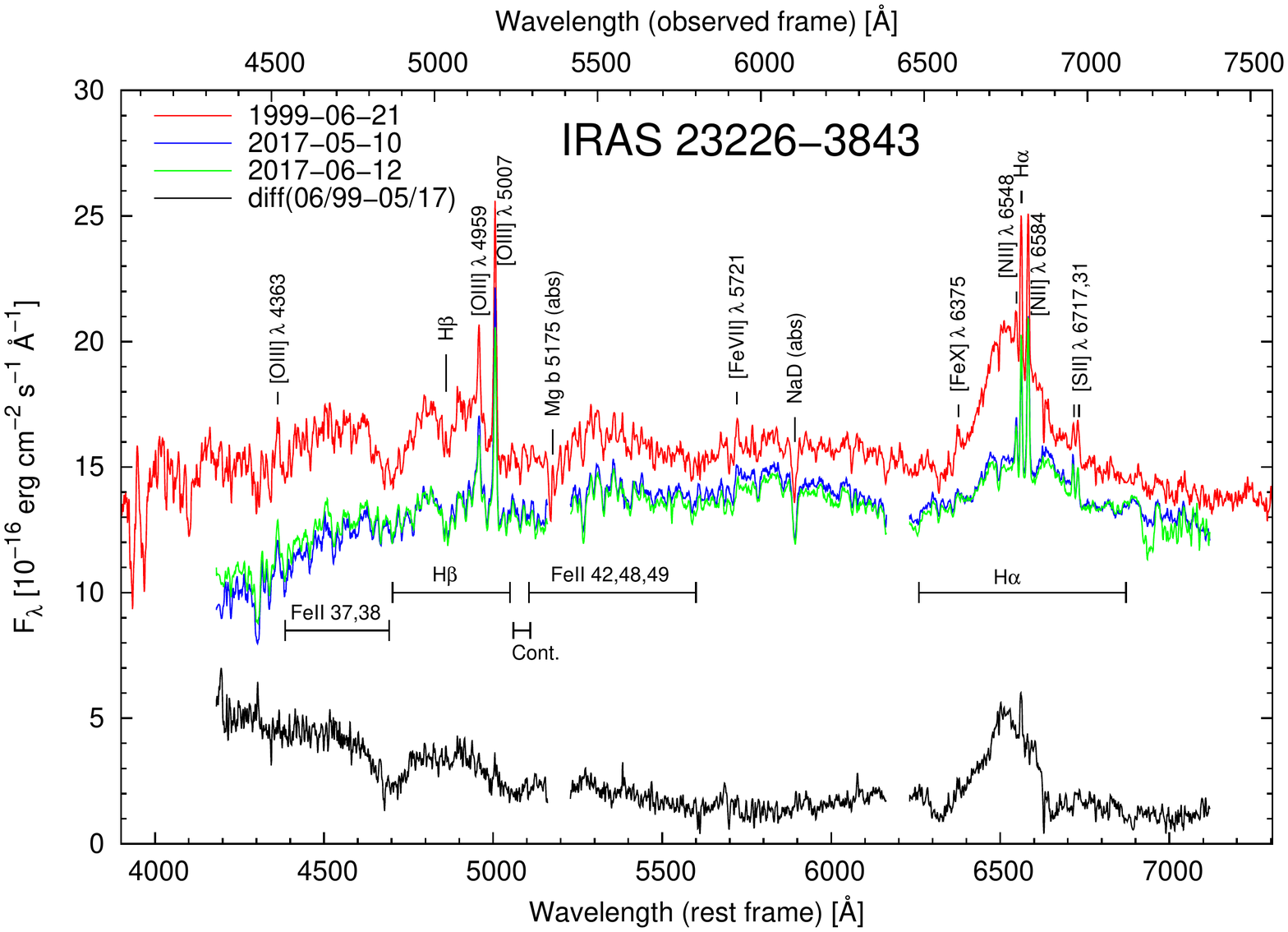}
\vspace*{-20mm} 
\caption{Optical spectra of 
\iras{} taken in 2017 as well as in
1999. The spectrum at the bottom gives the difference between the spectra taken in 1999
and May, 2017.}
\label{EH_IRAS23226spectra_all.pdf}
\end{figure*}

We show in Figure~\ref{EH_IRAS23226spectra_all.pdf}
the optical spectra of \iras{} that were taken with the SALT telescope
in May and June 2017.  
In addition, we present for comparison the optical spectrum taken with the 4m Blanco telescope at
the Cerro Tololo Observatory on 1999 June 21 
(Grupe et al.\citealt{grupe04}).
Furthermore, we show the difference spectrum
between 1999 and May 2017 when \iras{} was in a minimum state.
\begin{figure*}
\centering
\includegraphics[width=18.cm,angle=0]{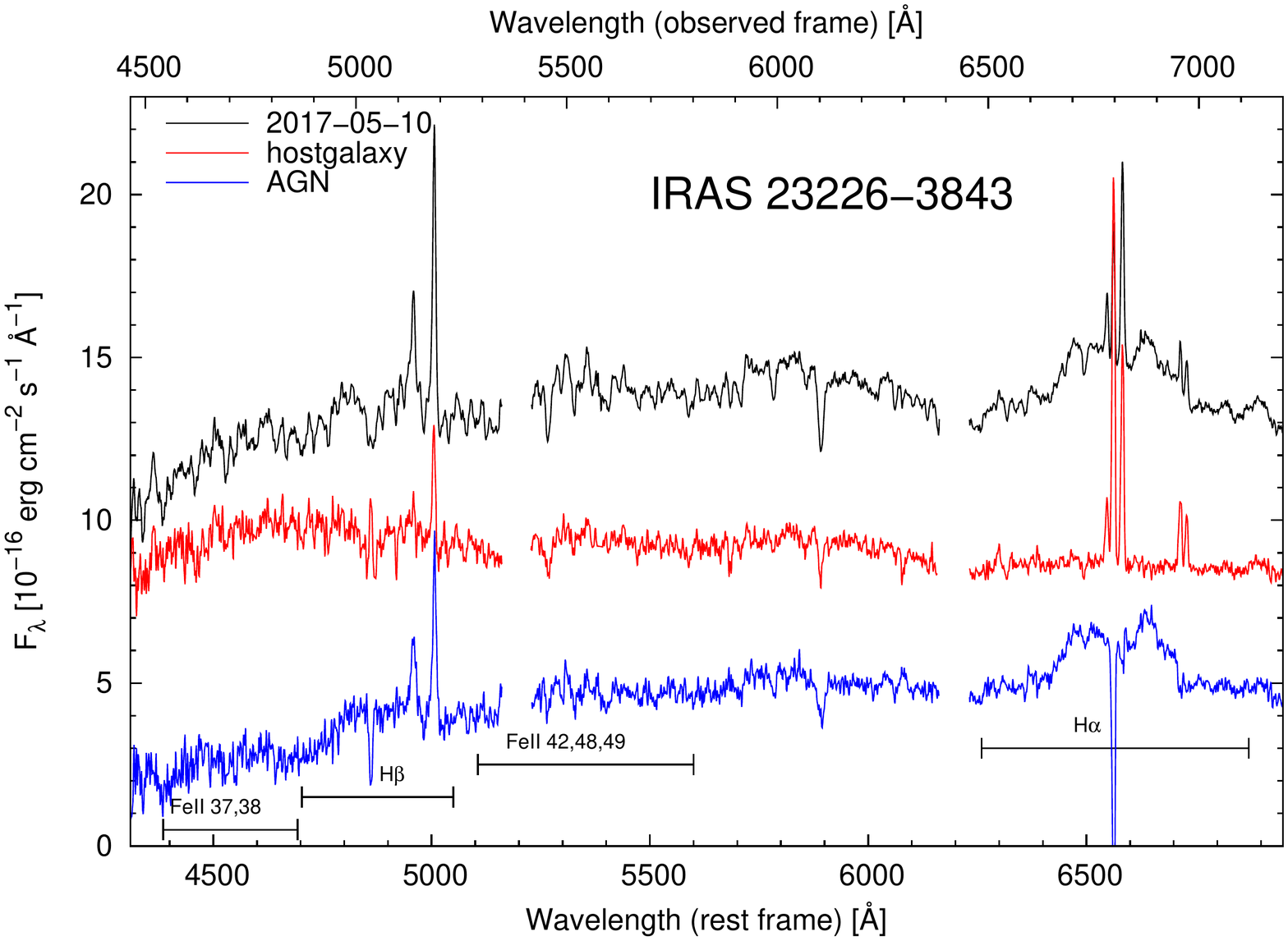}
\vspace*{-10mm} 
\caption{Central optical spectrum of 
\iras{} and a spectrum of the host galaxy (June 2017).
The spectrum at the bottom gives the broad lines and the nonthermal continuum component
after subtraction of the host galaxy spectrum.}
\label{EH_IRAShost_subtr.pdf}
\end{figure*}
The original optical spectra were taken with different telescopes and different 
spectrographs and also under different
atmospheric conditions. Therefore we had to intercalibrate our spectra.
The intercalibration of the spectra in Figure~\ref{EH_IRAS23226spectra_all.pdf}
was carried out with respect to the [\ion{O}{iii}]\,$\lambda$5007 line as well as
with respect to the narrow
H$\alpha$, [NII], and [SII] lines next to H$\alpha$ because the flux of the narrow emission lines
is considered to be constant on timescales of years.
 Because the spectrum of \iras{}  shows a strong additional contribution from the 
 underlying host galaxy and the narrow emission line region is extended, a relative flux accuracy 
 of only about 5\% was achieved for the nuclear emission line spectra.

The spectrum taken in 1999 exhibits an asymmetric broad
H$\alpha$ emission, a broad double-peaked H$\beta$ emission, and permitted broad FeII blends
of the multiplets 37 and 38  and 42, 48, 49. 
These broad H$\alpha$ and H$\beta$ lines as well as FeII blends 
are variable, as can be seen 
in the difference spectrum between 1999
and 2017 (see the bottom spectrum in Figure~\ref{EH_IRAS23226spectra_all.pdf}).
In addition to the broad lines, the typical narrow AGN emission lines of [OIII], [OII], [NII], [SII],
a narrow H$\alpha$, and the coronal lines [FeVII]5721 and [FeX]6375 are present.
The contribution of the host galaxy spectrum is strong in this AGN
based on its overall spectral behavior and on the strong stellar  
absorption features.

The wavelength ranges of the continua and pseudo-continua 
we used for the broad Balmer lines and FeII blends
are given in Table~\ref{wavelength-ranges}.
The pseudo-continua were used to subtract a continuum below
the variable broad emission lines.
\begin{table}
\caption{Wavelength ranges of the (pseudo-) continua, 
the Balmer lines, and FeII blends.}
\label{wavelength-ranges}
\begin{tabular}{lcc}
\hline
Continuum, Line & Wavelength range \\
           &    [\AA]         \\
\hline
Cont 4385    & 4375 -- 4390   \\
Cont 5085    & 5060 -- 5110    \\
Cont 5640    & 5620 -- 5660   \\
Cont 6120    & 6095 -- 6145    \\
Cont 6880    & 6870 -- 6890    \\
\hline
FeII(37,38)  & 4390 -- 4700       \\
H$\beta_{broad}$ & 4700 -- 5050   \\  
FeII(42,48,49)   & 5110 -- 5600   \\
H$\alpha_{broad}$ & 6260 -- 6870  \\  
\hline
\end{tabular}
\end{table}
We present in Table~\ref{NEL-intensities} the rest frame emission-line intensities based on
the spectrum taken in 1999.
\begin{table}
\caption{Optical emission line fluxes in \iras{} and line fluxes relative
to H$\beta$ in 1999.}
\label{NEL-intensities}
\begin{tabular}{lcc}
\hline
Line & Flux & Relative Fluxes\\
\hline
$\left[\text{OIII}\right]$4363               & 23.3$\pm$2.    & 0.038 \\ 
FeII(37,38)                         & 467.3 $\pm$30. & 0.78 \\
H$\beta_{broad}$                    & 600.7$\pm$25.  & 1. \\
$\left[\text{OIII}\right]$4959               & 27.9$\pm$2.    & 0.046 \\ 
$\left[\text{OIII}\right]$5007               & 92.7$\pm$3.    & 0.15 \\ 
FeII(42,48,49)                      & 362.0$\pm$30.  & 0.60 \\
$\left[\text{FeVII}\right]$5721              & 13.2$\pm$1.    & 0.022 \\ 
$\left[\text{FeX}\right]$6375                & 9.3$\pm$1.     & 0.015 \\ 
$\left[\text{NII}\right]$6548                & 12.9$\pm$1.    & 0.021 \\ 
H$\alpha_{narrow}$                  & 44.7$\pm$2.    & 0.074 \\ 
H$\alpha_{broad}$                   & 1192.6$\pm$40. & 1.98 \\   
$\left[\text{NII}\right]$6584                & 53.1$\pm$2.    & 0.088 \\ 
$\left[\text{SII}\right]$6717                & 12.5$\pm$1.    & 0.021 \\ 
$\left[\text{SII}\right]$6731                & 16.1$\pm$1.    & 0.027 \\ 
\hline
\end{tabular}
\tablefoot{
Line fluxes in units 10$^{-16}$\,erg\,s$^{-1}$\,cm$^{-2}$
}
\end{table}
We corrected the derived FeII($\lambda\lambda$5110-5600\AA) flux for the Mgb absorption.
Furthermore, we list in Table~\ref{BLR_variability} the broad-line
fluxes of H$\alpha$, H$\beta$, and of the FeII blends (37 and 38) and (42, 48, and 49)
for the three observing epochs in 1999 and 2017. In addition, we present
the observed continuum flux at 5085$\pm$25\AA{} as well as the
continuum flux after subtraction of the host galaxy continuum flux of
9.19 x 10$^{-16}$\,erg\,s$^{-1}$\,cm$^{-2}$\,\AA$^{-1}$.
We also give the continuum and line flux ratios with respect to the spectrum taken
in 1999.
\begin{table*}
\caption{Intensities of the optical continuum and of the integrated broad lines with
their relative variations.}
\label{BLR_variability}
\begin{tabular}{l c c c c c c}
\hline
date & Cont & Cont$_{w.o.host}$ & FeII(37,39) & H$\beta_{broad}$ & FeII(42,48,49) & H$\alpha_{broad}$\\ 
\hline
1999-6  & 15.27$\pm$0.4 & 6.08$\pm$0.4 & 467.$\pm$30. & 601.$\pm$25. & 362.$\pm$30. & 1193.$\pm$40. \\ 
2017-5  & 13.19$\pm$0.3 & 4.00$\pm$0.3 & 360.$\pm$30. & 316.$\pm$25. & 316.$\pm$30. & 629.$\pm$30. \\  
2017-6  & 12.94$\pm$0.3 & 3.75$\pm$0.3 & 391.$\pm$30. & 338.$\pm$25. & 339.$\pm$30. & 570.$\pm$30. \\ 
\hline
99/17-5 &   1.16       &    1.52      &  1.30  &    1.90      &   1.15       &  1.90  \\   
99/17-6 &   1.18       &    1.62      &  1.19  &    1.78      &   1.07       &  2.09  \\ 
\hline
\end{tabular}
\tablefoot{
Continuum flux in units of 10$^{-16}$\,erg\,s$^{-1}$\,cm$^{-2}$\,\AA$^{-1}$,\\
Line fluxes in units 10$^{-16}$\,erg\,s$^{-1}$\,cm$^{-2}$.
}
\end{table*}

The host galaxy contribution in the optical spectra of \iras{} is strong,
as has been reported before.
Based on the spectrum taken in May 2017, we derived a mean host galaxy 
spectrum by extracting a spectrum between 1.9\arcsec and 3.2\arcsec on both sides of the nucleus.  
This mean host galaxy spectrum is free of
broad Balmer lines and FeII blends.
The spectra of the central region (AGN + host galaxy),
of the host galaxy, and of the clean AGN spectrum
after subtraction and correction
for the underlying host galaxy component
are shown in Figure~\ref{EH_IRAShost_subtr.pdf}.
We scaled the host galaxy spectrum in such a way that the stellar absorption features
were minimized in the nuclear difference spectrum.
Finally, we subtracted the scaled host galaxy spectrum from the nuclear spectrum.
The final difference spectrum is composed of the central nonthermal component, the
broad emission lines, and some narrow [OIII] emission line residuals. 
These narrow emission lines
originate not only in the nucleus, but also come from off-nuclear regions probably
as a result of starbursts and/or shock heating. 

The variations of the optical continua and of the integrated broad lines
are shown in Figure~\ref{EH_IRAS23226spectra_all.pdf} and Table~\ref{BLR_variability}.
In addition, we present the continuum and line flux ratio of the observations in 1999
(high state) with respect to the fluxes in 2017 (low state) in Table~\ref{BLR_variability}.
The clean continuum after subtraction of the host galaxy component was fainter  by a factor of 1.5 in 2017
than in 1999.
The broad H$\alpha$ and H$\beta$ emission lines fluxes decreased by a factor of 1.9
and the FeII blends only decreased by a factor of 1.2.
The continuum and Balmer line fluxes of the spectra from May and June 2017 are
 identical to within 5\%.
The equivalent width (EW) of the [OIII]$\lambda$5007 line amounts to 14.6$\pm$2.\AA{}
(log[OIII]=1.16$\pm$.05)
in 1999 and to 18.8$\pm$2.\AA{} (log[OIII]=1.27$\pm$.05) in 2017 after correction for
the host galaxy contribution.

\subsubsection{\bf Balmer emission line profiles and their variations}

Based on the spectra presented in
Figs.~\ref{EH_IRAS23226spectra_all.pdf} and \ref{EH_IRAShost_subtr.pdf}, it is evident
that the Balmer lines in \iras{} are very broad. 
We present in Table~\ref{Balmerlinewidths} 
the line width FWHM and the full width at zero intensity (FWZI)
of H$\alpha$ and H$\beta$. 
\begin{table*}
\caption{Balmer line width FWHM and FWZI (left wing, right wing, total) in units of {\kms}
for 1999 and May 2017}
\label{Balmerlinewidths}
\begin{tabular}{l c c c c c c}
\hline
line (date) & FWHM(left) & FWHM(right) & FWHM & FWZI(left) & FWZI(right) & FWZI\\ 
\hline
H$\beta$\,(99) & -6120$\pm$200 & 6960$\pm$200 &  13080$\pm$300 & -9770$\pm$500 & 9720$\pm$500 & 19500$\pm$700 \\
H$\alpha$\,(99)& -5200$\pm$500 & 3150$\pm$500 &  8360$\pm$700  & -9780$\pm$500 & 10650$\pm$500 & 20430$\pm$700 \\ 
H$\beta$\,(17) & -5490$\pm$200 & 6930$\pm$200 &  12420$\pm$300 & -8550$\pm$800 & 9800$\pm$500 & 18350$\pm$1000 \\
H$\alpha$\,(17)& -5150$\pm$200 & 5920$\pm$200 &  11060$\pm$300 & -9340$\pm$800 & 8400$\pm$800 & 17740$\pm$1000 \\
\hline
\end{tabular}
\end{table*}
The FWHM of H$\beta$ and  H$\alpha$ amounts to
$\sim$12\,000\,km\,s$^{-1}$ in all spectra, except for the H$\alpha$ profile in 1999,
which showed a strong asymmetry.
The FWZI in all spectra amounts to $\sim$19\,000\,km\,s$^{-1}$.

\begin{figure}
\centering
\includegraphics[width=9.0cm,angle=0]{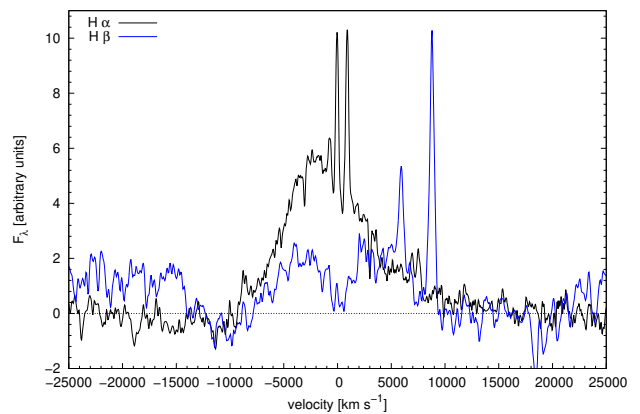}
\caption{Line profiles of H$\alpha$ and H$\beta$ (for 1999) in velocity space
after subtraction of the host galaxy spectrum.}
\label{EH_veloplot.pdf}
\end{figure}

We present in Figure~\ref{EH_veloplot.pdf} a comparison of the
H$\alpha$ and H$\beta$ line profiles in velocity space
after subtraction of the underlying continua for the spectrum taken in 1999.
The comparison of the H$\alpha$ and H$\beta$ line profiles shows considerable differences
between the profiles. 
The H$\alpha$ profile is asymmetric with respect to v=0~\kms{}. 
More specifically, the center of the broad component is blueshifted
by -1\,880\,km\,s$^{-1}$ with respect to 
the narrow H$\alpha$ component. 
The H$\beta$ profile, however, is symmetric with respect to v=0~\kms{} and shows a
double-peaked profile. It is evident that in the high state, the 
Balmer decrement H$\alpha$/H$\beta$ 
is different for the red and blue wings.

\begin{figure}
\centering
\includegraphics[width=9.cm,angle=0]{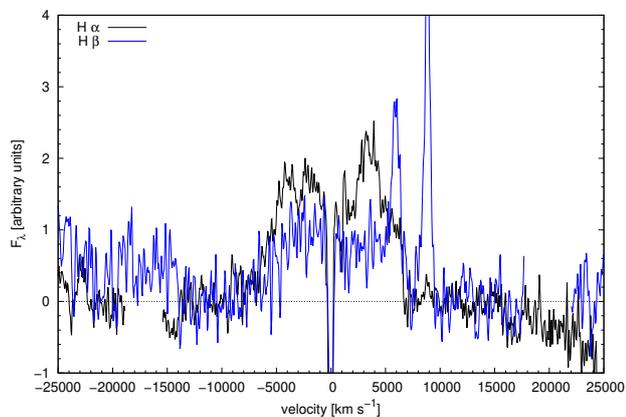}
\caption{Line profiles of H$\alpha$ and H$\beta$ (for May 2017) in velocity space
after subtraction of the host galaxy spectrum.}
\label{EH_veloplotN2.pdf}
\end{figure}

We present in Figure~\ref{EH_veloplotN2.pdf} a comparison of the
H$\alpha$ and H$\beta$ line profiles in velocity space for May 2017,
when \iras{} was in the low state.
As mentioned before, we first subtracted a spectrum of the host galaxy as
seen in Figure~\ref{EH_IRAShost_subtr.pdf} because the relative contribution
of the host galaxy is very strong. 
The subtraction of the host galaxy spectrum
leads to a central absorption in H$\alpha$ and H$\beta$
because the Balmer emission is extended in \iras{}.
During the low state, 
the profiles of both the broad H$\alpha$ and H$\beta$ line are symmetric.
Their double-peaked shape is typical for an underlying accretion disk. 
The line profile variations in the red and blue wings of H$\alpha$ and H$\beta$
differ from 1999 to 2017:
H$\alpha$ shows considerably stronger variations in the blue wing than in the red wing.
The Balmer decrement (1.8  -- 2.0)  is quite low and different for
1999 and 2017.

\section{Discussion}

\subsection{\bf X-ray, UV, and optical continuum variations}

The X-ray, UV, and optical continuum light curves of \iras{}  show the same
 variability trend from 2007 to 2017 
 (Figures~\ref{fig_dirk_xrt_uvot_lc07_17.pdf}
and \ref{fig_dirk_xrt_uvot_lc2017.pdf}).
The decreasing trend during 2017 was strongest in the X-rays
and less pronounced in the optical B and V bands (Fig.~\ref{fig_dirk_xrt_uvot_lc2017.pdf}).
\iras{} decreased in the X-rays by a factor of more than 30
from 2007 until 2017
(Swift, ROSAT). The variability amplitudes become
systematically smaller from the X-ray to the optical \swift{} bands
from a factor of 35 to a factor of about 2 (see Table~\ref{swiftvarstatistics}).
The U, B, V \swift{} bands were not corrected for the contribution of the host galaxy.
The broad H$\alpha$ line flux also decreased
by a factor of 2 from 1999 until 2017.

\begin{figure}
\centering
\includegraphics[width=10.cm,angle=0]{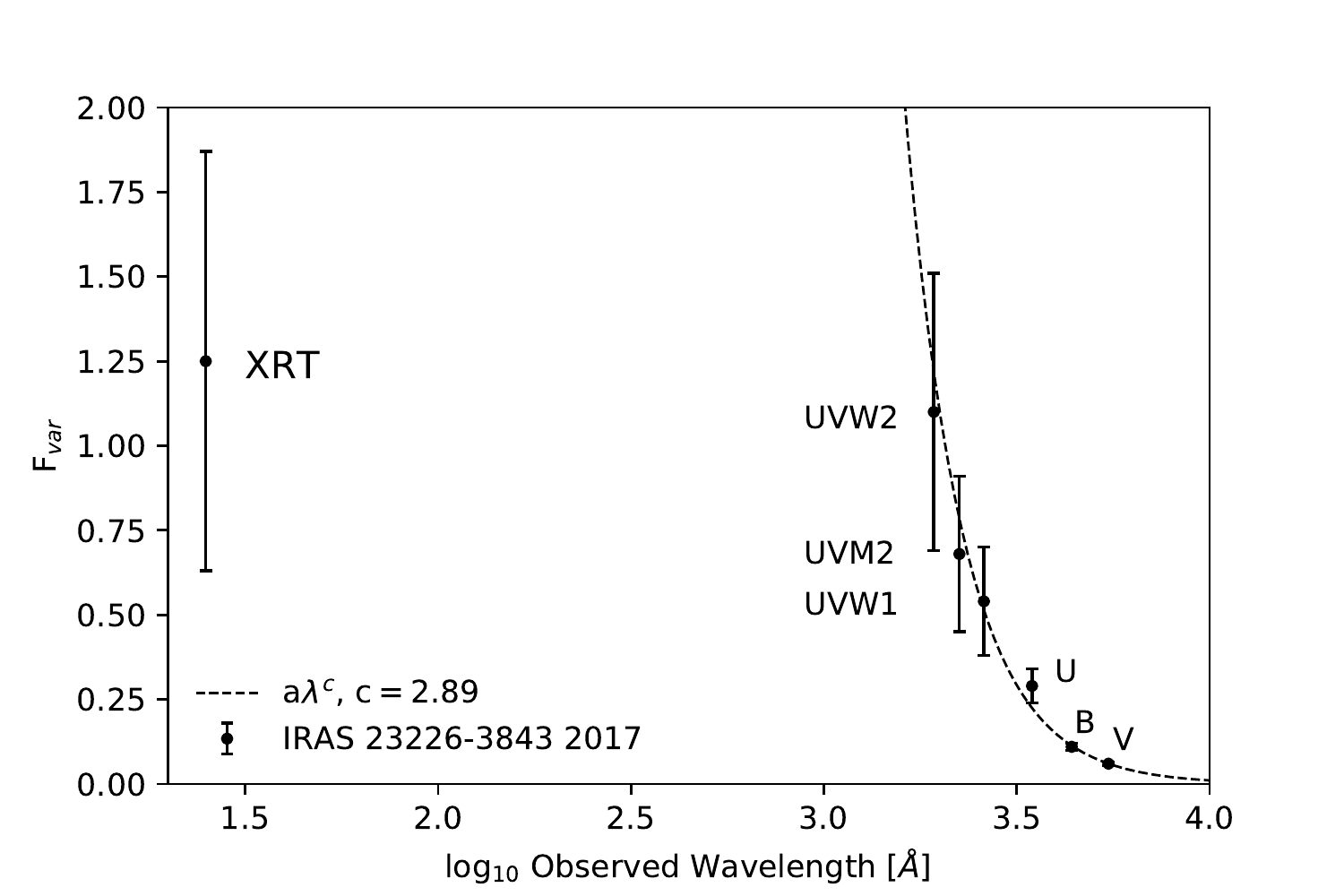}
\caption{Fractional variations of the X-ray, UV, and optical continuum bands measured from the 
\swift{} data as a function of wavelength.}
\label{OchmF_var.pdf}
\end{figure}

The fractional variation F$_{\rm var}$ is another way to describe the variability strength
in AGN. 
We present the fractional variation of the X-ray band together with the fractional variations
of the UV and optical bands in Table~\ref{swiftvarstatistics} and Figure~\ref{OchmF_var.pdf}.
The fractional variations are stronger at shorter wavelengths.
Interestingly, the
fractional variation in X-rays does not follow the same trend as
the fractional variations in the UV and optical bands,
and the extrapolation of the fit in the UV and optical bands
does not agree with the X-ray observations.
This is an indication that the origin of the UV/optical emission is
not a simple extension of the origin of the X-ray continuum emission.
A similar behavior has been found  in the changing-look AGN HE\,1136-2304
(Zetzl et al.\citealt{zetzl18}).
The prototype AGN NGC\,5548 shows an even lower correlation of the X-rays
with the UV/optical continuum (Edelson et al.\citealt{edelson15}).

We now compare the fractional variations in the UV and optical bands in  \iras{} with those
of the variability campaigns in HE\,1136-2304 and NGC\,5548 (Zetzl et al.\citealt{zetzl18},
Edelson et al.\citealt{edelson15},  Fausnaugh et al.\citealt{fausnaugh16}):
The fractional variations in the UV and optical bands in the changing-look HE\,1136-2304
were stronger by a factor of 2.3 than in NGC\,5548.
The fractional variations in our present study of \iras{} are even stronger by a factor of 2
 than those in HE\,1136-2304, indicating the extreme flux variations in  \iras{}.

\subsection{\bf X-ray spectrum}

The broadband \xmm/\nustar\ spectrum is power-law dominated with a contribution
from photoionized emission from cold gas, likely the outer accretion disk or torus.
The fit is improved when a small amount of warm absorption at low energies is included,
but this is not clearly resolved by the EPIC instruments and therefore might instead be due to
a more complex continuum or soft excess. The RGS data do not have
enough signal to determine how much absorption is present because the source flux is low.

There is a possible weak absorption line at $\sim7.3$~keV, with a residual feature
in both \xmm\ and \nustar\ spectra. If this were genuine, it would indicate a mildly
relativistic outflow, with a velocity of 0.05$c$. However, this
feature is not statistically significant ($\sim2\sigma$), therefore we cannot claim a detection.
The spectrum shows no evidence for a significant column of neutral absorption,
which can cause X-ray low states in some sources (e.g., Parker et al. 2014), 
or a relativistic reflection component, which can dominate when
the primary continuum drops strongly (e.g., Fabian et al. 2012).

\subsection{\bf Optical spectrum: extremely broad and double-peaked Balmer profiles}
 
\iras{} was classified as Seyfert 1 type
 (Allen et al.\citealt{allen91}) based on a spectrum taken between 1985 and 1990.  
 It was of clear Seyfert type 1 in 1999 as well (see Figure~\ref{EH_IRAS23226spectra_all.pdf}).  
The broad H$\alpha$ line profile was asymmetric. 
The center of the broad H$\alpha$
component was blueshifted with respect to the narrow emission by
-1900 \kms{}  (see  Figure~\ref{EH_veloplot.pdf}). 
In contrast to this, H$\beta$ shows a broad symmetric double-peaked 
emission line profile. 

It is generally accepted that broad double-peaked profiles originate
in relativistic Keplerian disks of gas surrounding the central supermassive black hole
(e.g., Eracleous \& Halpern \citealt{eracleous94}). 
During the minimum state in 2017, the two Balmer lines
(H$\alpha$ and H$\beta$) show the same symmetric double-peaked profile.
The peak separation corresponds to 7000 \kms{}. 
The broad Balmer line components, especially of H$\alpha,$ are faint and barely visible in 2017
in comparison to the optical spectra taken in 1999.
\iras{} was of Seyfert type 1.9 during the minimum state in 2017.
When the optical spectra taken in May and June 2017
are compared (Figure~\ref{EH_IRAS23226spectra_all.pdf}), the broad 
H$\alpha$ component was slightly stronger in May 2017.

The Balmer decrement H$\alpha$/H$\beta$ had a very low value of lower than 2 in 2017 
and only about 2 in 1999 (see Table~\ref{BLR_variability}). These values are
far below the expected
Case B recombination value of 2.7 (Osterbrock \& Ferland\citealt{osterbrock06}).
The observed unusual flat Balmer decrement in \iras{} indicates a very high hydrogen density of
n$_{H} > $ 10$^{11}$\,cm$^{-3}$ to the center of the accretion disk
(see the discussion of intrinsic hydrogen-line ratios in Gaskell,\citealt{gaskell17}).

The asymmetry of H$\alpha$  during the maximum state in 1999 might have been caused by absorption
of the red component. 
Partial dust obscuration of the BLR by outflowing dust clumps, a model introduced 
by Gaskell \& Harrington (\citealt{gaskell18}), can produce asymmetries
and velocity-dependent lags.
However, the derived Balmer decrement in \iras{} is far below the Case B value of 2.7 and 
therefore does not indicate extinction. The X-ray spectrum was unabsorbed as well
during the minimum state. The observed low H$\alpha$/H$\beta$ ratios in the emission line wings
are rather an indicator for high densities in the innermost clouds.
It was previously discovered by Shuder\cite{shuder82}, for instance, that the Balmer decrement
is flatter in the outer emission line wings that originate closer to the center.
During the high state of \iras{} in 1999, other cloud regions might have been ionized
than in the conditions in 2017.

There are clear signs for an additional blue absorption component in the H$\alpha$ and H$\beta$
line profiles at v=\,-11\,000\,km\,s$^{-1}$ 
(Figures~\ref{EH_IRAS23226spectra_all.pdf} and \ref{EH_veloplot.pdf}).
Intriguingly, there is further evidence for a possible weak absorption line at $\sim7.3$~keV in X-rays,
indicating  a mildly relativistic outflow  with v=\,-11\,000\,km\,s$^{-1}$. 
A blue absorption component in the Balmer lines at v$\sim$\,-11\,000\,km\,s$^{-1}$ 
like this has also been detected in another very broad-line AGN (Mrk\,926)
(Kollatschny\citealt{kollatschny10}), indicating a high-velocity outflow component.

The broad double-peaked Balmer line profiles and their strong variations
in \iras{}  are 
similar to those of other broad double-peaked galaxies such as NGC\,1097
(Storchi-Bergmann et al.\citealt{storchi97}), or broad-line radio galaxies
such as 3C390.3 (Shapovalova et al.\citealt{shapovalova10}) and Arp102B
(Shapovalova et al.\citealt{shapovalova13}).
Strateva et al. \cite{strateva03} studied the properties
of double-peaked Balmer line AGN in a large sample of SDSS galaxies.
In comparison to all these double-peaked AGN, \iras{} shows significantly 
stronger FeII line blends.
In general, broad-line AGN are expected to show only weak FeII line blends 
according to Eigenvector 1 studies of AGN
(Boroson \& Green\citealt{boroson92}, Sulentic et al.\citealt{sulentic00}). The Eigenvector 1 is
the dominant trend in AGN in which many properties correlate 
with the relative strength of optical FeII and [O III] line emission/equivalent widths.
Not a single AGN in the sample of about 20,000 SDSS quasars 
(Shen \& Ho\citealt{shen14}, their Fig.1)
shows such broad  H$\beta$ line 
widths of 12,400\,km\,s$^{-1}$ (in 2017) or 13.000\,km\,s$^{-1}$ (in 1999) 
in the two-dimensional Eigenvector 1 plane.

We  determined relative FeII strengths
EW(FeII)/EW(H$\beta$) of 0.6 (in 1999) and 1.0 (in 2017)
in \iras{}.
These values agree with the derived [OIII]$\lambda$5007 equivalent widths of 
14.6$\pm$2.\AA{} (1999) and 18.8$\pm$2.\AA{} (2017) in the two-dimensional Eigenvector 1 plane.
However, the position of \iras{} is unique in the two-dimensional Eigenvector 1 plane 
of Shen \& Ho\citealt{shen14} based on its extreme broad Balmer line widths 
in combination with relative strong FeII emission.
One way to explain this discrepancy between \iras{} and the SDSS quasar sample
might be the fact that the flat Balmer decrement and therefore the high-density emission line region
is unique in \iras{} in comparison to the SDSS quasars 
(Dong et al.\citealt{dong08}, Gaskell\citealt{gaskell17}).
In addition, \iras{} 
is only a low-luminosity Seyfert galaxy 
with a low Eddington ratio in contrast
to the more luminous SDSS quasars in the sample of Shen \& Ho\cite{shen14}.
In addition, a low Eddington ratio of L/L$_{edd}$=0.01 has been derived before by Grupe et al.\cite{grupe04}
for \iras{} during the high state
based on its broad emission lines and therefore high black hole mass of
$M_\text{BH}=1.7\times10^{8}M_{\sun}$.
The flat X-ray spectral slope $\alpha_{x}$=0.60  (Grupe et al.\citealt{grupe10})
is in accordance with a low Eddington ratio in \iras{}. 
Low L/L$_{edd}$ supports a disk-wind model for the broad-line region (MacLeod et al.\citealt{macleod19}).

\subsection{\bf Changing-look characteristics in \iras{}}

The broad-line profiles in \iras{} changed from a Seyfert 1 to a Seyfert 1.9 type
during the observing period from 1999 until 2017.
The general decrease in the broad-line components and in the optical continuum 
by a factor of 2.5 from 1999
to 2017 in combination with the change of the Seyfert type is accompanied by a strong
decrease of the
\swift{} X-ray flux by a factor of 35. These are the characteristics of a changing-look AGN. 
The spectral type variations  followed
the continuum intensity variations on
timescales of months to years. 

A similar spectral type scenario with decreasing continuum flux as seen in \iras{} 
has been discovered before in Fairall 9.
In the case of Fairall\,9, the optical continuum flux even dropped  to 20\%\ of its original
flux level within six years, and the spectral type changed
 from a quasar/Seyfert 1 type to a Seyfert 1.95 type
 (Kollatschny \& Fricke\citealt{kollatschny85}).
More recently, many new
 cases of changing-look sources have been identified in AGN in single
  well-studied nearby AGN (e.g., Parker et al.\citealt{parker19}, Oknyansky
  et al.\citealt{oknyansky19},
  Trakhtenbrot et al.\citealt{trakhtenbrot19}, 
   Wang et al.\citealt{wang19})  and in dedicated searches of large databases (e.g.,
  Graham et al.\citealt{graham20}).
 Spectral variations of the Seyfert type generally occur on timescales of years, such as in the changing-look AGN HE\,1136-2304  (Kollatschny et al.\citealt{kollatschny18},  
and references therein).
The case of IRAS 23226-3843 is special in comparison to HE\,1136-2304 and other changing-look AGN because its X-ray drop is extremely
high and its Balmer-line profile is exceptionally broad
and double-peaked.  

A change of the luminosity is always connected with a change of the Edddington ratio
 L/L$_{edd}$ in AGN because the central black hole mass does not change on short timescales. 
\iras{} showed a low Eddington ratio of L/L$_{edd}$=0.01
in its high state in 1999.
The optical luminosity was even lower by a factor of 2.5  for the low state in 2017.
The observed low Eddington ratio in \iras{} matches investigations of Noda \& Done\cite{noda18} 
and MacLeod et al.\cite{macleod19}, for example.
Noda \& Done\cite{noda18} suggested that all changing-look AGNs are associated with 
the state transition at Eddington ratios of a few percent. 
In a similar manner, extreme variable quasars show systematically lower
Eddington ratios (Rumbaugh et al.\citealt{rumbaugh18}).
MacLeod et al.\cite{macleod19} noted that changing-look quasars are at lower Eddingtin ratios 
than the overall quasar population in the SDSS catalog. 

Based on the observed long-term light curve, a tidal disruption event or microlensing can be excluded
as an explanation for the variability pattern in \iras{}.
Absorption-based variability is unlikely because the Balmer lines show no
indication for large-scale optical extinction and no X-ray absorption could be proved during the
minimum state. 
Various physical scenarios and their timescales
to explain the changing-look phenomenon are discussed in Stern et al.\cite{stern18}, for instance: 
The relevant timescales for changes at the inner accretion 
disk surrounding the central black hole are either the thermal or the heating/cooling front timescale. 
It has been pointed out by Lawrence\cite{lawrence18}, for example that the viscous timescale 
of a viscous radial inflow is too long in comparison to the observed timescales of months to years.
Most likely triggers for the observed changes on timescales of years in \iras{} 
are therefore magnetorotational instabilities (e.g., Ross et al.\citealt{ross18}) 
or accretion disk instabilities (e.g., Nicastro\citealt{nicastro03}).

\section{Summary}

We presented results of an optical, UV, and X-ray variability study of \iras{}
with  XMM-Newton, NuSTAR, \swift{}, and SALT. Our findings are summarized below.

\begin{enumerate}[(1)]
\item  The optical, UV, and X-ray continuum light curves showed the same
 variability pattern for 1999 until 2017.

\item  There were extreme X-ray, UV, and optical variability amplitudes in \iras{}.
 It varied in the X-ray continuum by a factor of 35 and in the optical by a factor of 2.
 
 \item The spectral type of \iras{}
changed from a clear Seyfert 1 type 
to a Seyfert 1.9 type, confirming its changing-look AGN character.

\item The broadband \xmm/\nustar\ spectrum is power-law dominated, with a contribution
from photoionized emission from cold gas, likely the outer accretion disk or torus.

\item We did not find evidence that the continuum and line variability (the changing look) 
 is driven by absorption: while there is tentative evidence for a remarkable outflow component at
 v=\,-11\,000\,km\,s$^{-1}$  seen in absorption, the broadband X-ray spectrum is unabsorbed
 and the optical broad Balmer lines do not show evidence for extinction. Therefore we favor
 true changes caused by accretion disk instabilities or magnetorotation instabilities
 as the driving mechanism behind the changing look. 

\item In addition to the broad Balmer lines, \iras{} exhibits strong FeII blends, 
in contrast to what is expected from Eigenvector 1 studies.
 
\item \iras{} shows exceptionally broad and double-peaked Balmer lines. 
The broad double-peaked profiles originate
in a relativistic Keplerian disk of gas surrounding the central supermassive black hole.
The unusually low Balmer decrement H$\alpha$/ H$\beta$ with a numerical value of two only  indicates a very high hydrogen density n$_{H} > $ 10$^{11}$\,cm$^{-3}$ to the center of the accretion disk.
 
\end{enumerate}

\begin{acknowledgements}
WK thanks Hartmut Winkler for lively debates on the optical spectra of \iras{}.
He also thanks Gary Ferland for valuable comments.
This paper is based on observations obtained with XMM-Newton, an ESA science mission with
instruments and contributions directly funded by ESA Member States and NASA.
Based on observations obtained with the Southern African Large Telescope.
We thank the \swift{} team for performing the ToO observations. 
This research has made use of the XRT Data Analysis Software (XRTDAS) developed
under the responsibility of the ASI Science Data Center (ASDC), Italy.
This research has made use of the NASA/IPAC Extragalactic
Database (NED) which is operated by the Jet Propulsion Laboratory,
Caltech, under contract with the National Aeronautics and Space
Administration. 
This work  made use of data from the NuSTAR mission, a project led by the
California Institute of Technology, managed by the Jet Propulsion
Laboratory, and funded by the National Aeronautics and Space Administration.
This research has made use of the NuSTAR Data Analysis Software (NuSTARDAS)
jointly developed by the ASI Science Data Center (ASDC, Italy) and the
California Institute of Technology (USA).
This work has been supported by the DFG grant
Ko 857/33-1.
\end{acknowledgements}

\clearpage

\begin{appendix}
\section{\bf Additional tables}
\begin{table*}
\caption{\label{swiftlog} 
 XRT and UVOT monitoring observation log: 
Julian date, UT date, and XRT and UVOT exposure times in seconds.
}
\begin{tabular*}{\textwidth}{@{\extracolsep{\fill} }ccrrrrrrr}
\hline 
\noalign{\smallskip}
Julian date &  \\
2\,400\,000+&  \rb{UT date (middle of the exposure)}  &  \rb{XRT} & \rb{V} & \rb{B} & \rb{U} & \rb{UV W1} & \rb{UV M2} & \rb{UVW2}   \\
\hline 
54266.6750 & 2007-06-15 04:12 &  5037 & --- & --- & --- & --- & --- & 4970 \\
54312.9792 & 2007-07-31 11:30 &  6213 & --- & --- & 6127 & --- & --- & --- \\
54361.9743 & 2007-09-18 11:23 & 15767 & --- & --- & --- & --- & --- & --- \\ 
54363.0389 & 2007-09-19 12:56 &  6645 & --- & --- & --- & --- & --- & --- \\ 
54364.1139 & 2007-09-20 14:44 &  8616 & --- & --- & --- & --- & --- & --- \\ 
54676.7875 & 2008-07-29 06:54 &  1095 & 104 & 104 & 104 & 208 & 114 & 417 \\
55092.9722 & 2009-09-18 11:20 & 17422 & --- & --- & --- & --- & --- & --- \\
56493.4166 & 2013-07-20 10:13 &  1461 & 119 & 119 & 119 & 237 & 340 & 475 \\
56556.3861 & 2013-09-20 21:16 &   522 &  44 &  44 &  44 &  87 & 114 & 176 \\
57578.0396 & 2016-07-08 12:57 &  6263 &  84 &  85 &  84 & 168 & 260 & 5509 \\
57856.3778 & 2017-04-12 21:04 &  1141 &  47 &  47 &  47 &  94 & 134 & 188 \\
57865.0097 & 2017-04-21 12:14 &  1111 &  84 &  84 &  84 & 169 & 295 & 337 \\
57880.7952 & 2017-05-07 07:05 &   936 &  78 &  78 &  78 & 154 & 222 & 309 \\
57891.6903 & 2017-05-18 04:34 &   629 &  48 &  48 &  48 &  97 & 180 & 195 \\
57895.1028 & 2017-05-21 14:28 &   946 &  74 &  74 &  74 & 150 & 222 & 301 \\
57896.0243 & 2017-05-23 00:35 &   962 &  80 &  80 &  80 & 158 & 234 & 317 \\
57902.8451 & 2017-05-29 08:17 &   979 &  80 &  80 &  80 & 160 & 245 & 321 \\
57909.5556 & 2017-06-05 01:20 &   202 &  --- &  31 &  31 &  64 & --- &  58 \\
57913.2708 & 2017-06-08 18:30 &   924 &  74 &  74 &  74 & 148 & 223 & 297 \\ 
57916.4201 & 2017-06-11 22:06 &  1898 & 136 & 136 & 136 & 510 & 401 & 544 \\
57923.4986 & 2017-06-18 23:59 &   941 &  26 &  79 &  79 & 158 & 210 & 317 \\
57931.2806 & 2017-06-26 18:44 &   445 &  37 &  37 &  37 &  76 &  90 & 151 \\
57938.1931 & 2017-07-03 16:37 &   799 &  64 &  64 &  64 & 127 & 192 & 257 \\
57944.5639 & 2017-07-10 01:32 &   832 &  82 &  82 &  82 & 163 &  63 & 326 \\
\hline 
\end{tabular*}
\label{swiftlog}
\end{table*}

\begin{sidewaystable*}
\centering
\tabcolsep1.mm
\caption{\swift{} monitoring: V, B, U, UVOT W1, M2, and W2 reddening-corrected
flux
 in units of 10$^{-15}$W\,m$^{-2}$(10$^{-12}$ ergs s$^{-1}$ cm$^{-2}$; columns 2
to 7)
and magnitudes in the Vega system (columns 8 to 13).
}
\begin{tabular}{lcccccc|cccccccc}
\hline
\noalign{\smallskip}
JD-2400000& V & B & U &UVW1  & UVM2 & UVW2  & {V} & {B} & {U} & {UV W1} & {UV
M2} & {UVW2}   \\
\noalign{\smallskip}
(1) & (2) & (3) & (4) & (5) & (6) & (7) & (8) & (9) & (10) & (11) & (12) & (13)
 \\  
\noalign{\smallskip}
\hline
\noalign{\smallskip}
54266.6750 &  --- & --- & --- & --- & --- & 7.34\plm0.33 &  --- & --- & --- &
--- & --- & 15.09\plm0.05 \\
54312.9792 &  --- & --- & 10.99\plm0.47 & --- & --- & --- &  --- & --- &
15.11\plm0.05 & --- & --- & --- \\
54361.9743 &  --- & --- & --- & --- & --- & ---  &  --- & --- & --- & --- & ---
& --- \\
54363.0389 &  --- & --- & --- & --- & --- & ---  &  --- & --- & --- & --- & ---
& --- \\
54364.1139 &  --- & --- & --- & --- & --- & ---  &  --- & --- & --- & --- & ---
& --- \\
54676.7875 &  21.59\plm1.00 & 17.18\plm0.73 & 11.74\plm0.63 & 9.79\plm0.61 &
10.14\plm0.67 & 8.45\plm 0.56 &  14.94\plm0.05 & 15.55\plm0.05 & 15.04\plm0.05 &
15.06\plm0.06 & 15.03\plm0.07 & 15.21\plm0.07 \\
55092.9722 &  --- & --- & --- & --- & --- & ---  &  --- & --- & --- & --- & ---
& --- \\
56493.4166 &  16.64\plm0.77 & 12.62\plm0.58 & 6.71\plm0.39 & 4.53\plm0.30 &
3.40\plm0.23 & 3.23\plm0.23  &    15.23\plm0.05 & 15.88\plm0.05 & 15.65\plm0.06
& 15.90\plm0.06 & 16.22\plm0.07 & 16.25\plm0.07 \\
56556.3861 & 18.82\plm1.15 & 14.27\plm0.78 & 7.54\plm0.51 & 5.44\plm0.27 &
5.33\plm0.49 & 4.23\plm0.33  &     15.09\plm0.07 & 15.75\plm0.06 & 15.52\plm0.07
& 15.70\plm0.08 & 15.89\plm0.10 & 15.97\plm0.09 \\
57578.0396 & 16.05\plm0.83 & 11.65\plm0.58 & 5.61\plm0.35 & 3.31\plm0.27 &
2.69\plm0.22 & 2.25\plm0.16 &      15.27\plm0.06 & 16.00\plm0.05 & 15.84\plm0.07
& 16.23\plm0.08 & 16.47\plm0.09 & 16.64\plm0.06 \\
57856.3778 & 16.85\plm0.84 & 11.21\plm0.68 & 5.14\plm0.40 & 2.51\plm0.25 &
2.53\plm0.27 & 2.10\plm0.21  &     15.22\plm0.07 & 16.01\plm0.07 & 15.94\plm0.08
& 16.54\plm0.11 & 16.54\plm0.12 & 16.72\plm0.10 \\
57865.0097 & 15.40\plm0.83 & 11.75\plm0.58 & 5.02\plm0.31 & 2.96\plm0.26 &
2.36\plm0.19 & 2.31\plm0.19  &     15.31\plm0.06 & 15.96\plm0.05 & 15.96\plm0.07
& 16.37\plm0.08 & 16.60\plm0.09 & 16.62\plm0.08 \\
57880.7952 & 16.17\plm0.88 & 11.94\plm0.58 & 5.77\plm0.35 & 2.78\plm0.24 &
2.30\plm0.21 & 2.16\plm0.19 &      15.26\plm0.06 & 15.94\plm0.05 & 15.81\plm0.07
& 16.43\plm0.08 & 16.64\plm0.10 & 16.69\plm0.09 \\
57891.6903 & 16.58\plm1.07 & 12.76\plm0.73 & 5.02\plm0.39 & 2.88\plm0.31 &
2.34\plm0.23 & 2.17\plm0.19  &     15.23\plm0.07 & 15.88\plm0.06 & 15.96\plm0.08
& 16.40\plm0.10 & 16.63\plm0.11 & 16.68\plm0.10 \\
57895.1028 & 16.93\plm0.94 & 10.87\plm0.60 & 5.06\plm0.31 & 2.34\plm0.28 &
2.38\plm0.22 & 1.98\plm0.17  &     15.21\plm0.06 & 16.04\plm0.06 & 15.95\plm0.07
& 16.62\plm0.10 & 16.61\plm0.10 & 16.79\plm0.09 \\
57896.0243 & 15.87\plm0.88 & 11.79\plm0.24 & 5.57\plm0.35 & 2.76\plm0.23 &
2.17\plm0.20 & 2.02\plm0.16  &     15.28\plm0.06 & 15.95\plm0.05 & 15.85\plm0.07
& 16.44\plm0.08 & 16.71\plm0.10 & 16.76\plm0.09 \\
57902.8451 & 16.93\plm0.88 & 11.70\plm0.58 & 5.06\plm0.31 & 2.63\plm0.23 &
2.66\plm0.22 & 1.94\plm0.16  &     15.21\plm0.06 & 15.96\plm0.05 & 15.95\plm0.07
& 16.50\plm0.09 & 16.49\plm0.09 & 16.81\plm0.09 \\
57909.5556 &  --- & 12.47\plm0.83 & 4.91\plm0.43 & 2.54\plm0.29 & --- &
1.72\plm0.26 & ---                  & 15.89\plm0.07 &  15.98\plm0.10 &
16.53\plm0.12 & --- & 16.94\plm0.17 \\
57913.2708 & 17.29\plm0.94 & 11.79\plm0.58 & 4.67\plm0.31 & 2.36\plm0.22 &
2.38\plm0.21 & 2.05\plm0.17  &     15.19\plm0.06 & 15.95\plm0.05 & 16.04\plm0.07
& 16.61\plm0.09 & 16.61\plm0.10 & 16.75\plm0.09 \\
57916.4201 & 16.05\plm0.76 & 11.21\plm0.48 & 5.14\plm0.27 & 2.67\plm0.17 &
2.29\plm0.16 & 1.97\plm0.15  &     15.27\plm0.05 & 16.01\plm0.05 & 15.93\plm0.06
& 16.48\plm0.07 & 16.65\plm0.08 & 16.79\plm0.08 \\
57923.4986 & 16.46\plm0.88 & 11.65\plm0.58 & 5.14\plm0.31 & 2.69\plm0.21 &
2.34\plm2.19 & 2.11\plm0.17  &     15.24\plm0.06 & 15.97\plm0.05 & 15.94\plm0.07
& 16.47\plm0.08 & 16.62\plm0.10 & 16.72\plm0.09 \\
57931.2806 & 16.05\plm0.48 & 11.99\plm0.73 & 5.38\plm0.43 & 2.66\plm0.30 &
2.23\plm0.30 & 2.10\plm0.22 &      15.26\plm0.08 & 15.94\plm0.07 & 15.89\plm0.09
& 16.48\plm0.11 & 16.68\plm0.15 & 16.72\plm0.11 \\
57938.1931 & 17.17\plm1.00 & 12.67\plm0.68 & 4.59\plm0.31 & 2.73\plm0.26 &
2.02\plm0.20 & 1.97\plm0.17 &      15.19\plm0.06 & 15.88\plm0.06 & 16.06\plm0.08
& 16.45\plm0.09 & 16.78\plm0.11 & 16.79\plm0.09 \\
57944.5639 &  15.64\plm0.83 & 12.09\plm0.58 & 4.67\plm0.33 & 2.66\plm0.25 &
1.84\plm0.22 & 2.02\plm0.16 &     15.29\plm0.06 & 15.93\plm0.05 & 16.04\plm0.07
& 16.48\plm0.08 & 16.89\plm0.19 & 16.66\plm0.08 \\
\noalign{\smallskip}
\hline                                                                          
                                     
\noalign{\smallskip}
\label{swiftmag}
\end{tabular}
\end{sidewaystable*}

\begin{table*}
\caption
{\swift{} monitoring: Julian date, UT date, XRT 0.3--10 keV count rates (CR),
and hardness ratios (HR$^1$), X-ray photon index $\Gamma$, the observed 
0.3--10 keV X-ray flux in units of $10^{-12}$ erg s$^{-1}$ cm$^{-2}$, and
 reduced $\chi^2$ of the simple power-law model fit (pl).
}
\tabcolsep+1mm
\begin{tabular*}{\textwidth}{@{\extracolsep{\fill} } lclccc}
\hline 
\noalign{\smallskip}
Julian Date &  \\
2\,400\,000+ & \rb{CR} & \,\quad\rb{HR} &    \rb{$\Gamma_{\rm pl}$} & \rb{XRT flux} & \rb{$(\chi^2/\nu)_{\rm pl}$} \\
\hline 
54266.6750 & 0.288\plm0.008 & +0.208\plm0.024 & 1.88\plm0.08 & 11.38\plm0.26 & 39.8/50 \\
54312.9792 & 0.215\plm0.006 & +0.178\plm0.028 & 1.96\plm0.08 &  8.29\plm0.31 & 47.0/49 \\
54361.9743 & 0.076\plm0.003 & +0.266\plm0.033 & 1.78\plm0.09 &  3.10\plm0.14 & 48.7/42 \\
54363.0389 & 0.079\plm0.004 & +0.386\plm0.051 & 1.67\plm0.14 &  3.51\plm0.25 & 11.4/18 \\
54364.1139 & 0.072\plm0.003 & +0.255\plm0.044 & 1.89\plm0.11 &  2.81\plm0.09 & 25.9/23 \\
54676.7875 & 0.268\plm0.017 & +0.311\plm0.055 & 1.70\plm0.17 & 11.45\plm0.07 & 121/152$^5$ \\ 
55092.9722 & 0.315\plm0.005 & +0.138\plm0.015 & 2.04\plm0.04 & 11.46\plm0.02 & 216.0/163 \\
56493.4166 & 0.067\plm0.007 & +0.305\plm0.096 & 1.82\plm0.29 &  2.91\plm0.35 & 65.1/74$^5$ \\
56556.3861 & 0.137\plm0.017 & +0.241\plm0.127 & 1.92\plm0.39 &  5.09\plm0.92 & 49.5/50$^5$ \\
57578.0396 & 0.048\plm0.003 & +0.345\plm0.056 & 1.77\plm0.16 &  2.10\plm0.24 & 149.4/172$^5$ \\
57856.3778 & 0.041\plm0.009 & +0.294\plm0.159 & 1.78\plm0.47 &  1.65\plm0.33 & 27.7/33$^5$ \\
57865.0097 & 0.035\plm0.007 & +0.057\plm0.177 & 2.04\plm0.62 & 1.35\plm0.40 & 19.0/24$^5$ \\
57880.7952 & 0.037\plm0.007 & +0.562\plm0.136 & 2.30\plm1.02 & 2.48\plm0.16 & 18.4/26$^5$ \\
57891.6903 & 0.035\plm0.011 & +0.383\plm0.232 & --- & 1.43\plm0.42$^6$ & --- \\
57895.1028 & 0.033\plm0.006 & +0.554\plm0.153 & 1.77\plm0.58 & 1.41\plm0.45 & 12.2/22$^5$ \\
57896.0243 & 0.024\plm0.006 & -0.224\plm0.241 & 2.25\plm0.67 & 1.96\plm0.34 & 16.8/16$^5$ \\
57902.8451 & 0.018\plm0.007 & +0.348\plm0.310 & --- & 0.67\plm0.27$^6$ & --- \\
57909.5556 & 0.017\plm0.013 & --- & --- & 0.65\plm0.50$^6$ & --- \\
57913.2708 & 0.026\plm0.006 & --0.002\plm0.220 & 2.52\plm0.66 & 1.30\plm0.32 & 14.0/16$^5$ \\
57916.4201 & 0.019\plm0.004 & +0.028\plm0.200 & 2.01\plm0.89 & 0.33\plm0.12 & 15.2/28$^5$ \\
57923.4986 & 0.016\plm0.005 & --0.211\plm0.310 & --- & 0.62\plm0.29$^6$ & --- \\
57931.2806 & 0.014\plm0.007 & --- & --- & 0.54\plm0.27$^6$ & --- \\
57938.1931 & 0.017\plm0.006 & +0.056\plm0.190 & --- & 0.65\plm0.23$^6$ & --- \\
57944.5639 & 0.020\plm0.006 & +0.435\plm0.221 & --- & 0.77\plm0.23$^6$ & --- \\
\hline 
\end{tabular*}
$^1$ The hardness ratio is defined as HR = $\frac{hard-soft}{hard+soft}$ , where {\it \textup{soft}} and {\it \textup{hard}} are the background-corrected counts in the 0.3--1.0 keV and 1.0--10.0 keV bands, respectively 

$^2$ The intrinsic $N_{\rm H, intr}$ is given in units of $10^{21}$ cm$^{-2}$

$^3$ No additional absorber required. The fit is consistent with  Galactic absorption alone. 

$^4$ These observations were performed in windowed timing mode. 

$^5$ Fit using Cash statistics (Cash\citealt{cash79})

$^6$ Extrapolated from the XRT count rates using the flux and count rate of the MJD 57865 observation
\label{swiftdata} 
\end{table*}

\end{appendix}

\end{document}